\documentclass[letterpaper,iop,Times,revtex4]{emulateapj}
\usepackage[pass,paperwidth=8.5in,paperheight=11in]{geometry}
 \usepackage{graphicx}
 \usepackage{amsmath}
\usepackage{natbib}
\usepackage{gensymb}
\usepackage{float}
\usepackage{hyperref}
\usepackage{xcolor}
\hypersetup{
    colorlinks,
    linkcolor={blue!75!black},
    citecolor={blue!75!black},
    urlcolor={blue!80!black}
}

\slugcomment{{\sc Accepted to ApJ:}, February 20, 2015} 
\journalinfo{{\sc ApJ}, 803:72, 2015 April 20} 

\begin{document}

\shorttitle{Rayleigh-Taylor Unstable Flames}
\shortauthors{Hicks}

\title{Rayleigh-Taylor Unstable Flames -- Fast or Faster?}


\author{E. P. Hicks\altaffilmark{1}}
\affil{Center for Interdisciplinary Exploration and Research in
Astrophysics (CIERA) and the Department of Physics and Astronomy, Northwestern
University, Evanston, IL 60208}
\email{eph2001@columbia.edu}


\begin{abstract}
Rayleigh-Taylor (RT) unstable flames play a key role in the explosions of Type Ia supernovae.
However, the dynamics of these flames is still not well-understood. RT unstable flames are affected
by both the RT instability of the flame front and by RT-generated turbulence. The coexistence of
these factors complicates the choice of flame speed subgrid models for full-star Type Ia
simulations.  Both processes can stretch and wrinkle the flame surface, increasing its area and,
therefore, the burning rate.  In past research, subgrid models have been based on either the RT
instability or turbulence setting the flame speed. We evaluate both models, checking their
assumptions and their ability to correctly predict the turbulent flame speed. Specifically, we
analyze a large parameter study of 3D direct numerical simulations of RT unstable model flames. This
study varies both the simulation domain width and the gravity in order to probe a wide range of
flame behaviors.  We show that RT unstable flames are different from traditional turbulent flames:
they are thinner, rather than thicker when turbulence is stronger.   We also show that none of
several different types of turbulent flame speed models accurately predicts measured flame speeds.
In addition, we find that the RT flame speed model only correctly predicts the measured flame speed
in a certain parameter regime. Finally, we propose that the formation of cusps may be the factor
causing the flame to propagate more quickly than predicted by the RT model.
\end{abstract}


\section{Introduction}
\label{cha:intro}

Type Ia supernovae are extremely bright stellar explosions that are not only fascinating in their
own right, but also play an important role in cosmological distance measurements
\citep{riess1998,perlmutter1999}.  Type Ia supernovae are thought to be white dwarf
stars that explosively burn their carbon and oxygen into heavier elements, including
${}^{56}\text{Ni}$.  The radioactive decay of the ${}^{56}\text{Ni}$, in turn, produces the light
seen as the Type Ia explosion. Recent observations of SN 2011fe support the white dwarf explosion
scenario \citep{nugent2011,bloom2012,brown2012} but there is still debate about whether the
Type Ia supernova progenitor is two merging white dwarfs (the double-degenerate scenario), for
example, see \citet{iben1984,webbink1984,hicken2007,yoon2007,raskin2012,piro2014}, or a single white
dwarf driven to explosion by material accreted from a companion star (the single-degenerate
scenario), for example, see \citet{whelan1973,nomoto1982a,iben1984,marietta2000}. In this paper, we
will focus on how the Rayleigh-Taylor instability affects thermonuclear burning in the
single-degenerate scenario.

In the single-degenerate scenario, a white dwarf accretes material from a companion star until its mass approaches the
Chandrasekhar  limit.  During this process, the white dwarf becomes more compact until, somehow, a thermonuclear runaway is
triggered.  Burning engulfs the star and it explodes.  In one scenario, thermonuclear burning is initially triggered in a
convective region near the center of the star and then propagates outward \citep{woosley2007a,nonaka2012}. The thermonuclear
burning is expected to take the form of a very thin front that initially propagates at subsonic speeds.  This is known as a
deflagration. It was once thought that the deflagration might be enough to trigger the observed supernova explosion, but it
has since been shown that a deflagration-only event does not produce an energetic-enough explosion and results in incorrect
spectra, with low-velocity carbon and oxygen components \citep{gamezo2003,gamezo2004}. If, however, the deflagration is sped
up until it becomes a self-sustaining, supersonic burning wave (a detonation), then a more realistic explosion is predicted.
This more realistic scenario is known as the deflagration-to-detonation transition (DDT) and forms the basis of many
different single-degenerate Type Ia explosion scenarios including the standard DDT
\citep{blinnikov1986,woosley1990,khokhlov1991,khokhlov1997,khokhlov1997a,gamezo2004,ropke2007}, pulsational detonations
\citep{khokhlov1991,arnett1994a,arnett1994b,hoflich1995,hoflich1996,bravo2006}, and gravitationally confined detonations
\citep{plewa2004,jordan2008,meakin2009,seitenzahl2009,jordan2012}. The cause of the detonation remains an open question;
traditionally, the Zel'dovich gradient mechanism has been invoked \citep{khokhlov1997,khokhlov1997a}, but
\citet{poludnenko2011} have recently identified other processes that can trigger unconfined detonations. Finally, it has been
shown that a detonation-only explosion produces too much nickel and iron and can be ruled out
\citep{arnett1969,khokhlov1993,filippenko1997,gamezo1999}.  Although it could occur in many ways, a DDT is necessary for a
realistic Type Ia explosion.

In single-degenerate explosion scenarios, the initial deflagration is Rayleigh-Taylor (RT) unstable
\citep{rayleigh1883,taylor1950} because dense fuel sits above lighter burnt ashes in the star's
gravitational field.  The RT instability affects the flame in two different ways: first, it
stretches the flame surface; second, the nonlinear evolution of this stretching process generates
turbulence behind the flame front, which back-reacts on the flame surface, wrinkling it further
\citep{V05,zhang2007,hicks2013}. Both stretching and wrinkling increase the surface area of the
flame, speeding it up.  As the flame speeds up, it may eventually undergo a DDT. The details of the
DDT, in particular, when and how the transition to detonation occurs, determine critical observables
such as nickel production \citep{gamezo2003,gamezo2004,gamezo2005,ropke2007,krueger2012,
seitenzahl2013}. This transition is still not understood, but one possibility, the Zel'dovich
gradient mechanism \citep{zeldovich1970}, depends critically on the details of the conditions
produced by the deflagration \citep{khokhlov1997,khokhlov1999,oran2007,ropke2007a,ropke2007}. 
Without a full understanding of RT unstable flames, the mechanism and final nickel yields of this
class of Type Ia supernovae models will remain uncertain.

Ideally, the propagation of RT-unstable flames and the DDT would be studied using full-star
simulations.  However, the separation of scales in the problem makes this unfeasible: the size of
the star (approximately Earth-sized) is much too large relative to the width of the flame ($10^{-4}$
to $10^{2}$ cm according to \citet{timmes1992}) to resolve both in the same simulation
\citep{oran2005}. Instead, full-star simulations must include a variety of subgrid models,
including, in particular, a subgrid model that gives the speed of the flame below certain scales. 
There are two basic types of subgrid model, and there has been a long debate about which of the two
is correct. Each model incorporates a different assumption about how RT-unstable flames should
behave. In one, the turbulent flame speed is set by the Rayleigh-Taylor instability.  In the other,
the interactions of turbulence with the flame front dictate the flame speed. The question at the
heart of this and prior research \citep{hicks2013} is whether both or either of these two
deflagration subgrid models is physically appropriate.

RT-type subgrid scale (RT-SGS) models
\citep{khokhlov1995,khokhlov1996,gamezo2003,gamezo2004,gamezo2005,
zhang2007,townsley2007,jordan2008} are based on the hypothesis that the RT stretching of the flame
front sets the turbulent flame speed.  In these models, the turbulent speed of the flame on an
unresolved scale $\Delta$ is given by the velocity $v_{RT}(\Delta) \propto \sqrt{g \, A \, \Delta}$
which is naturally associated with the Rayleigh-Taylor instability at the length scale $\ell =
\Delta$.  Here $g$ is the gravitational acceleration and the Atwood number is $A = (\rho_{\rm
fuel}-\rho_{\rm ash})/(\rho_{\rm fuel}+\rho_{\rm ash})$, where $\rho_{\rm fuel}$ and $\rho_{\rm
ash}$ are the densities of the fuel and the ash.  Two major hypotheses underlie the RT-type subgrid
model: self-similarity and self-regulation.  Self-similarity means that the flame is effectively a
fractal, so the RT subgrid model applies at any scale.  Self-regulation means that physical
processes will force the flame back towards the RT flame speed if the flame starts to move too fast
or too slow. Self-regulation is a competition between two processes, the creation of flame surface
area by the RT instability (which increases the turbulent flame speed) and destruction of flame
surface area by cusp burning (which decreases the turbulent flame speed); cusps are areas of the
flame surface with high curvature.  As the flame develops small wrinkles, due to turbulence or the
RT instability, cusp burning ensures that these wrinkles will be destroyed, returning the flame
speed to the RT predicted value. Likewise, if wrinkle destruction is too effective, the flame front
becomes flatter and the RT instability more efficiently increases the surface area and the flame
speed.  The net result is that the flame is forced to travel at the RT value.

On the other hand, turbulence-based subgrid scale (Turb-SGS) models are based on the hypothesis
that flame behavior is determined by the interaction between turbulence and the flame front
\citep{niemeyer1995,niemeyer1997,niemeyer1997a,reinecke1999,ropke2005,
schmidt2006a,schmidt2006b,jackson2014}.  These models do not distinguish between different sources
of turbulence or whether the turbulence is upstream or downstream of the flame front.  Turb-SGS
models are adapted from the field of turbulent premixed combustion, which studies
the propagation of premixed flames (in various configurations) through pre-existing turbulence. In
these models, the turbulent flame speed is often based on the root-mean-square (rms) velocity of the
pre-existing, upstream turbulence.  The key assumption behind astrophysical Turb-SGS models is that
flames interact with upstream and downstream turbulence in the same way. One purpose of this paper
is to test that assumption.

An exploding white dwarf has two potential sources of turbulence:
turbulence produced by the convection that precedes ignition and turbulence produced by the
RT-unstable flame front.  If the pre-ignition core convection is strongly turbulent, then the flame
will travel through this pre-existing turbulence.  In that case, the flame is forced to interact
with every turbulent eddy it encounters as it propagates upstream.  This is exactly the case studied
by traditional turbulent premixed combustion, so models from that field are good candidates for
Turb-SGS models for Type Ia simulations. The second source of turbulence is the RT instability of
the flame front. As the RT instability deforms the flame front, the flame front produces
turbulence baroclinically. Previous studies have shown that this turbulence exists only downstream
of the flame front \citep{V03,V05,schmidt2006b,hicks2013}. The flame will not necessarily interact
with this turbulence because it does not need to travel through the turbulent region in order to
propagate upstream.  In this case, Turb-SGS models based on ideas from traditional turbulent
combustion may not apply because the physical situation is fundamentally different.  We will not
address the question of whether substantial pre-existing turbulence exists in the white dwarf,
see \citet{zingale2009,nonaka2012}.

The only way to determine which, or even whether, either type of subgrid model is correct is to
directly study RT-unstable flames.  There have been many such studies, which can be
organized by various criteria including dimensionality, resolution requirements, flame type,
evolution time and flame regime. 2D simulations
\citep{bell2004b,V03,V05,zhang2007,biferale2011,hicks2013} are less computationally expensive than
3D simulations and can cover a wider range of parameter space, but do not produce realistic
turbulence. 3D simulations
\citep{khokhlov1994,khokhlov1995,zingale2005a,zhang2007,ciaraldi-schoolmann2009,chertkov2009} treat
the turbulence correctly, but are more computationally expensive.  Simulations also differ in what
scale is resolved: some use a subgrid model themselves \citep{ciaraldi-schoolmann2009}, others
resolve the Gibson scale and the flame width \citep{bell2004b,zingale2005a}, and still others
resolve down to the viscous scale \citep{V03,V05,chertkov2009,hicks2013}.   RT-unstable flame
studies also use different treatments for the flame itself, from realistic carbon-oxygen flames
\citep{bell2004b,zingale2005a,ciaraldi-schoolmann2009} to thickened flames in a degenerate setting
\citep{zhang2007} to model flames in a Boussinesq setting \citep{V03,V05,chertkov2009,hicks2013}.
Carbon-oxygen flames are most realistic and directly applicable to supernovae, but model flames can
better isolate specific effects, such as RT stretching.  Another difference between studies is
whether they focus on the early, transient stages of RT-unstable flame growth
\citep{bell2004b,zingale2005a,zhang2007,chertkov2009} or later, saturated stages when the flame
speed varies around a statistically steady average \citep{V03,V05,zhang2007,hicks2013}.  In
simulations, the saturated stage is reached when the RT instability can no longer grow horizontally
due to confinement by the sides of the simulation domain.  In this case, a balance develops between
RT growth, which creates surface area, and burning, which destroys it. It is likely that RT flame
propagation in the star is not statistically steady because there is no confinement mechanism for RT
modes and the star expands as the flame propagates; however, it is still not known whether
unconfined flames can saturate.  So, which choice is more physically relevant -- statistically
unsteady or saturated simulations -- remains unclear. Even if the flame behavior is only transient
in the star, saturated simulations indicate the statistically steady state the flame is approaching,
even if it never reaches it.  The effect of boundary conditions on simulated RT unstable flames have
been specifically studied by \citet{V03,V05,hicks2014}.  Finally, simulations vary in what
parameter values they use and which combustion regime they probe: flamelets
\citep{bell2004b,zingale2005a,V03,V05,zhang2007,hicks2013}, thin reaction zones
\citep{bell2004b,zingale2005a,chertkov2009} or broken reaction zones \citep{chertkov2009}.

Other facets of burning in white dwarfs have been addressed in other types of studies.  If the
turbulence generated by the initial convective stage in the white dwarf is strong, then the flame
may be dominated by its propagation through this pre-existing turbulence instead of by the RT
instability or by the turbulence produced by the RT instability.  In that case, traditional ideas
and studies of turbulent combustion would be clearly applicable to the formulation of subgrid
models.  A small selection of the applicable papers, some with specific reference to the Type
Ia problem, include: \citet{aspden2008,aspden2010,poludnenko2010,poludnenko2011, poludnenko2011a,
hamlington2011}; \citet*{aspden2011}; \citet{hamlington2012, chatakonda2013}.
Even if the flame does not move into a strongly convective field, the turbulence from the carbon
flame could influence the trailing oxygen flame; this scenario has been studied
by \citet*{woosley2011} and \citet*{aspden2011c}.  Finally, it is likely that, after ignition
takes place near the core of the white dwarf, subsequent burning may take the form of rising
buoyant plumes or bubbles (the surfaces of which would be RT unstable).  The dynamics of these
plumes has been studied by \citet*{vladimirova2007,zingale2007} and \citet*{aspden2011a}.

In this paper, we will test the basic predictions of RT-SGS and Turb-SGS models against a large
parameter study of 3D, fully-resolved, RT unstable model flames.  To date, there have been few 3D
simulations of RT unstable flames
\citep{khokhlov1994,khokhlov1995,zingale2005,zhang2007,chertkov2009}, and no parameter studies large
enough to clearly test the scaling laws predicted by the subgrid models.  In particular, this is the
first set of 3D model flame simulations in the flamelet regime that fully resolve the viscous scale.
Resolving the viscous scale accounts for all possible interactions between the flame and turbulence
in the simulations (for a similar 2D study see \citet{hicks2013}).    The set of eleven simulations
discussed in this paper tests the scaling laws over a wide range of flame behavior, from a steady
rising bubble to a flame highly disturbed by the RT instability.  In particular, we will focus on
flames in the flamelet regime, a regime in which Type Ia flames are expected to spend a considerable
fraction of their time. In addition, we will look for a transition from the flamelets regime to the
reaction zones regime, as predicted by traditional turbulent combustion theory.  This transition is
important because it could lead to conditions that may cause a detonation.

In order to isolate the effects of the RT instability on the flame front, we made as many
simplifications to our parameter study setup as possible.  In doing this, we neglected many of the
complexities of real white dwarf flames. For example,  we used a simple model reaction instead of a
full chemical reaction chain.  We used the Boussinesq approximation and therefore ignored
compressibility effects and sound waves. These simplifications allow us to focus directly on the
effect that gravity has on the flame without having to disentangle it from other effects like the
Landau-Darrieus instability. Finally, we focused on the saturated state, in which quantities such as
the flame speed vary around a statistically steady average in order to obtain robust scalings that
don't depend on time.

In this paper, we will test the predictions of the two types of subgrid models both indirectly and
directly. To start, in Section \ref{sec:problem}, we describe the problem formulation, the control
parameters that are varied in the parameter study and provide a list of the simulations.  Next, in
Section \ref{sec:flameregimes}, we discuss the different combustion regimes predicted by traditional
turbulent combustion theory, and compare the predictions of this theory with observations from our
simulations. In particular, we show that the flame remains in the flamelets regime after it is
predicted to transition to the reaction zones regime and that the flame becomes thinner instead of
thicker when turbulence is strong. Then, in Section \ref{sec:speedtests}, we test the predictions of
both types of subgrid models, beginning with three types of turbulence-based subgrid models and
ending with the predictions of the Rayleigh-Taylor subgrid model.  After showing that all of these
models fail in certain regions of parameter space, we will (in Subsection \ref{ssec:cusps}) consider
the possibility that the formation of cusps by turbulence and/or the Rayleigh-Taylor instability
might explain these deviations. Finally, we draw some conclusions in Section
\ref{sec:theend}.

\section{Problem Formulation}
\label{sec:problem}

To isolate the effects of the Rayleigh-Taylor instability on the flame front, we simulated a simple
model flame.  This model makes two major simplifications to more realistic treatments of
nuclear burning: one simplification to the fluid equations themselves, and one simplification to the
treatment of the reaction.  To simplify the fluid equations, we employ the Boussinesq approximation,
which reduces the fully compressible Navier-Stokes equations to an incompressible form.  To simplify
the reaction, we use a simple model reaction which avoids the intricacies of a full reaction chain.

\begin{figure*}
\begin{center}
\includegraphics[height=7in,angle=0]{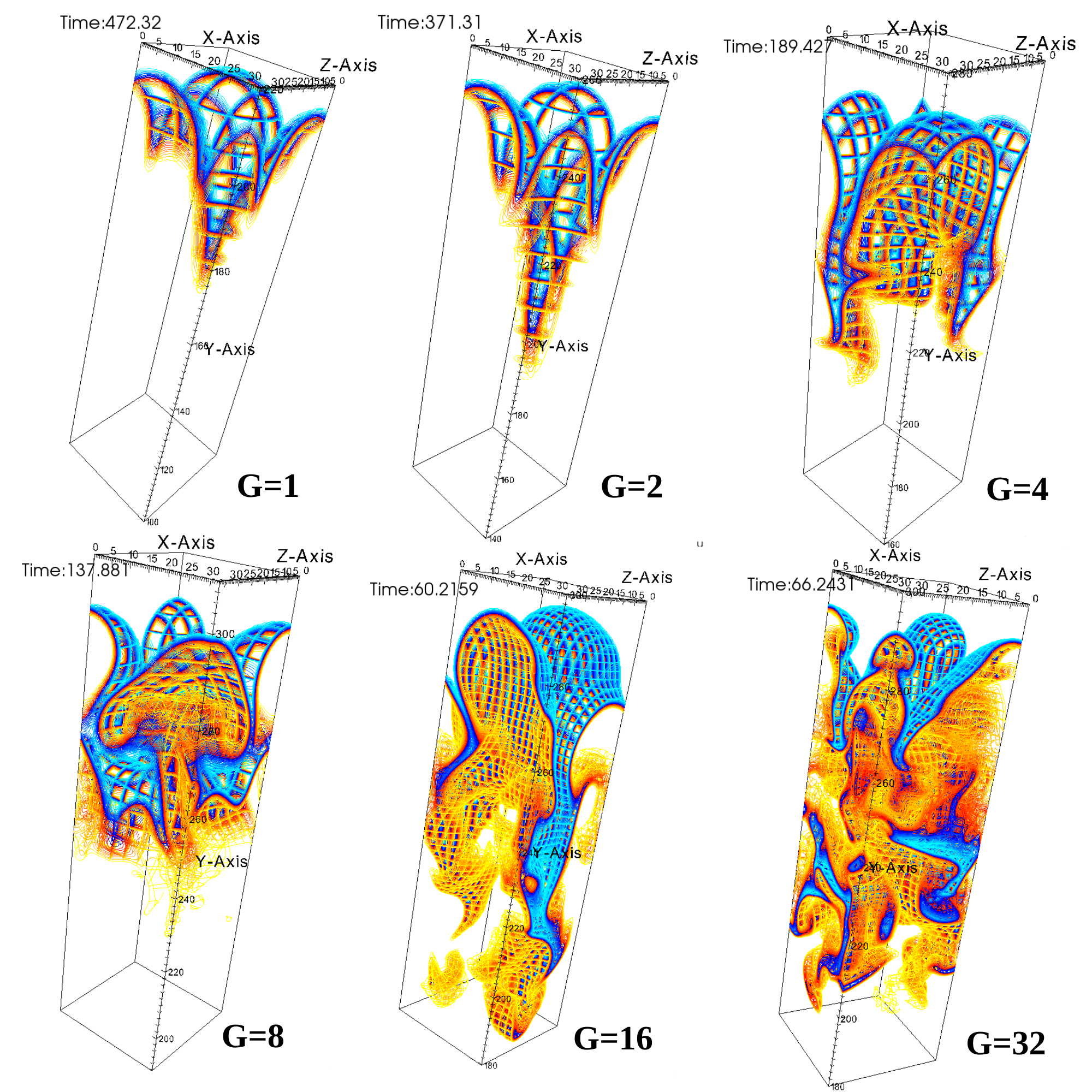}
\caption{Contour Plots of Temperature, L=32.  This figure shows a sample contour plot of temperature
for each of the six simulations with domain size $L=32$. Blue colors represent mostly unburnt fuel
and red/yellow colors represent mostly burned ashes.  A colorbar showing the assignment of colors to
temperatures is shown in Figure \ref{fig-L64-temps}.  Each flame is propagating in the
$y$-direction, against the force of gravity which points in the $-y$ direction.  In general, the
flame surface shape changes with time, causing the flame speed to vary. Note that the flame shape
ranges from a simple rising bubble at $G=1$ to a complex, highly convoluted surface at $G=32$.
Simulation A, discussed in Section \ref{ssec:cusps}, is the $G=16$ plot in this figure.}
\label{fig-L32-temps}
\end{center}
\end{figure*}

\begin{figure*}
\begin{center}
\includegraphics[height=7in,angle=0]{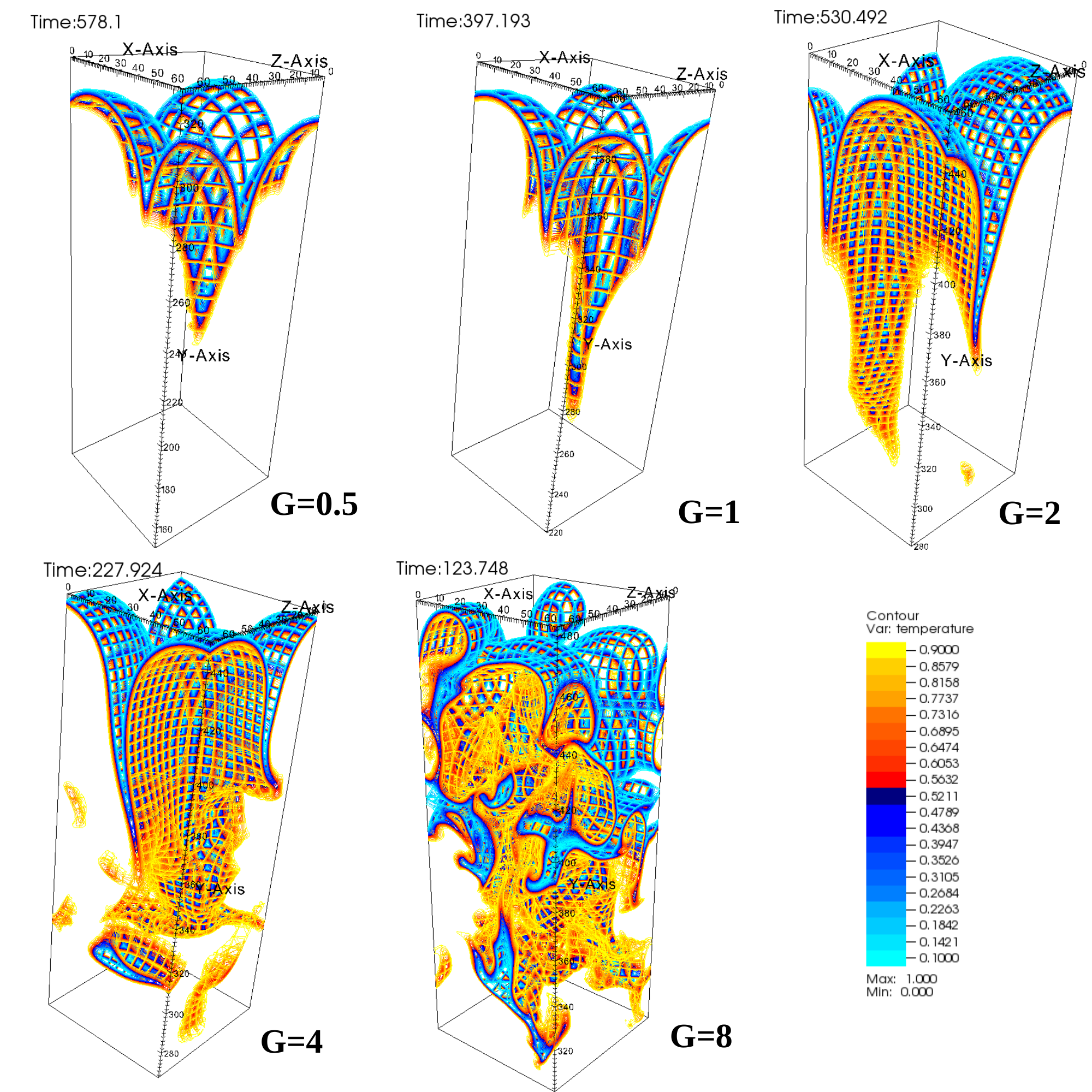}
\caption{Contour Plots of Temperature, L=64.  This figure shows a sample contour plot of temperature
for each of the five simulations with domain size $L=64$. Note that, similarly to Figure
\ref{fig-L32-temps}, the flame shape ranges from a simple rising bubble at $G=0.5$ to a complex,
highly convoluted surface at $G=8$. Simulation B, discussed in Section \ref{ssec:cusps}, is the
$G=8$ plot in this figure.}
\label{fig-L64-temps}
\end{center}
\end{figure*}

The Boussinesq approximation is appropriate for subsonic flows with only small density (and
temperature) variations and a small vertical extent compared to the scale height of the system
\citep{spiegel1960}.  If these criteria are fulfilled then a simplified set of equations can be
derived, in which the density differences in the flow are taken into account only in the
gravity-dependent buoyancy forcing term in the Navier-Stokes equation.  For combustion, the density
across the flame front $\Delta \rho$ is only included in the forcing term; all other terms depend
only on the density of the unburnt fuel, $\rho_o$.  In this approximation, the continuity equation
is incompressible.  The Boussinesq approximation disallows shocks and heating due to the viscous
dissipation of energy. A flame front may be Rayleigh-Taylor unstable (because gravitational forcing
due to density variations in the flow is accounted for in the buoyancy term) but can not be
Landau-Darrieus unstable (because density variations are not accounted for outside of the buoyancy
term). All of these simplifications are desirable so that the RT instability can be considered
without other complications.

The second simplification is that we added a simple reaction term, $R(T)$, to the
advection-diffusion-reaction (ADR) temperature equation to replace all of the details of
realistic nuclear burning.  In this model, $T$ is a reaction progress variable that tracks the state
of the fluid from unburnt fuel at $T=0$ to burnt ashes at $T=1$. The reaction progress
variable represents both the mass fraction of the burned material and the fraction of energy
released into the flow \citep{vladimirova2006}. In using this model, we do not consider any
specific chain of nuclear reactions or the separate evolution of nuclear species and temperature.
Instead, the ADR equation models both temperature and species evolution. This simplified approach
has been used by many other combustion studies and possible choices for $R(T)$ include the
Kolmogorov-Petrovkii-Piskunov (KPP), mth-order Fisher, bistable, Arrhenius and ignition reactions (a
review of model reaction types is given by \citet{xin2000}). In this study we chose $R(T) = 2 \alpha
T^2 (1-T)$, a bistable reaction with an ignition temperature of zero which, therefore, has no
bistable behavior. We choose this particular reaction instead of the more physically realistic
Arrhenius reaction because the reaction front is wider and therefore easier to resolve (see also
\citet{hicks2013}). We did not choose the KPP reaction used by \citet{V03,V05} because the KPP flame
front is very wide and the KPP reaction has an unstable fixed point at $T=0$ which makes it more
numerically unstable.

\begin{deluxetable*}{lllllllll}[tH]
\tablecaption{Simulation Parameters \label{table-sims} }
\tablewidth{0pt}
\tablehead{ \colhead{$G$} & \colhead{$L$} & \colhead{Physical Size} & \colhead{Elements}
& \colhead{Order} & \colhead{DOF} & \colhead{Resolution} & \colhead{Time} & \colhead{Time
Step ($10^{-3}$)}}
\startdata
      1  & 32 & 32 x 512 x 32 & 4 x 64 x 4    & 8  & 524288    & 1.000 & 504.06 & 30     \\
      2  & 32 & 32 x 576 x 32 & 4 x 72 x 4    & 10 & 1152000   & 0.800 & 429.63 & 64.801 \\
      4  & 32 & 32 x 576 x 32 & 8 x 144 x 8   & 7  & 3161088   & 0.571 & 250.22 & 18.051 \\ 
      8  & 32 & 32 x 608 x 32 & 8 x 152 x 8   & 11 & 12947968  & 0.364 & 157.71 & 6.659  \\ 
      16 & 32 & 32 x 608 x 32 & 16 x 304 x 16 & 7  & 26693632  & 0.286 & 75.561 & 3.665  \\
      32 & 32 & 32 x 640 x 32 & 16 x 320 x 16 & 9  & 59719680  & 0.222 & 87.16 & 1.99   \\ 

      0.5 & 64 & 64 x 640 x 64 & 8 x 80 x 8    & 8  & 2621440  & 1.000 & 705.03 & 15     \\ 
      1   & 64 & 64 x 704 x 64 & 8 x 88 x 8    & 8  & 2883584  & 1.000 & 440.92 & 65.071 \\ 
      2   & 64 & 64 x 768 x 64 & 16 x 192 x 16 & 7  & 16859136 & 0.571 & 606.63 & 13.031   \\ 
      4   & 64 & 64 x 832 x 64 & 16 x 208 x 16 & 9  & 38817792 & 0.444 & 300.39 & 6.924   \\
      8   & 64 & 64 x 832 x 64 & 16 x 208 x 16 & 11 & 70873088 & 0.364 & 173.48 & 4.316    \\ 
\enddata
\tablecomments{Simulation Parameters.  The columns are: the nondimensional gravity, the
nondimensional domain size, the physical size, the number of elements ($N_x$ x $N_y$ x $N_z$), the
polynomial order ($p_o$), the number of degrees of freedom ($ \sim N_x N_y N_z p_o^3$), the average
resolution (the average spacing between collocation points), the total running time, the
time step.  All quantities are in nondimensional units.}
\end{deluxetable*}

The bistable reaction has a simple, laminar solution in a stationary, gravity-free fluid
\citep{constantin2003}. When the flame is laminar, it is planar with a characteristic width of
$\delta$ and it travels with the laminar flame speed $s_o$. $\delta$ and $s_o$ are set by $\alpha$,
the laminar reaction rate, and $\kappa$, the thermal diffusivity, such that $s_o = \sqrt{\alpha
\kappa}$ and $\delta=\sqrt{\dfrac{\kappa}{\alpha}}$. The actual flame thickness ($\delta_t$), also
called the thermal flame width, is larger than the characteristic flame width ($\delta$) by a factor
of 4 ($\delta_t=4\delta$) as calculated by measuring the distance between the level sets $T=0.1$ and
$T=0.9$.    Finally, $\delta_i$ is the width of the flame reaction zone, the part of the flame in which
the most intense burning takes place.  This is typically 2-10 times smaller than the laminar flame width.

The fluid equations were non-dimensionalized by the characteristic length scale (the laminar flame
front thickness, $\delta$) and time scale in the problem (the reaction time, $1/\alpha$) \citep{V03}
to give
\begin{subequations}
\begin{gather}
 \dfrac{D \textbf{u}}{Dt}=-\left(\dfrac{1}{\rho_o}\right)\nabla p + G\,T + Pr
\nabla^{2} \textbf{u} \label{NNS}\\
\nabla \cdot \textbf{u}=0  \label{Nincomp}  \\
\dfrac{D T}{D t} = \nabla^{2} T + 2T^{2}(1-T). \label{Nheat}
 \end{gather}
\end{subequations}
with two control parameters:
\begin{align}
G &=g\left(\frac{\Delta \rho}{\rho_o}\right)\dfrac{\delta}{s_o^{2}} \\
Pr &=\frac{\nu}{\kappa}
\end{align}
where $G$ is the non-dimensionalized gravity and $Pr$ is the Prandtl number.  $G$ is positive if the
flame is moving in the opposite direction from the gravitational force, as is the case in these
simulations and in the white dwarf.  Here, $\rho_o$ is the density of the unburnt fuel and $\Delta
\rho$ is the increase in density across the flame front, so that $\rho(T) = \rho_o + \Delta \rho \,
T$. In this formulation, $p$ is the pressure deviation from hydrostatic equilibrium.  For
simplicity, $\nu$ (the kinematic viscosity) and $\kappa$, are taken to be constants independent of
temperature. The non-dimensional domain width, $L=\dfrac{\ell}{\delta}$, where $\ell$ is the
dimensional length in the $x$ and $z$ directions, is the third control parameter.  These
parameters can be translated into the densimetric Froude number, $Fr_d=\dfrac{1}{\sqrt{G L}}$. 
Another parameter that will be considered is the Reynolds number $Re = u' L$ (when $Pr=1$) which is
calculated from the root-mean-square (rms) velocity measured in the flow (see Section
\ref{ssec:measure}). Finally, the Lewis number, $Le = \kappa / D$ (where $D$ is the material
diffusivity), is effectively $Le=1$ because the simulations only track temperature and do not
separately consider material diffusivity.  In the simulations presented in this paper, $G$ and $L$
are varied but $Pr=1$.

All simulations used Nek5000 \citep{nek5000}, a freely-available, open-source, highly-scalable
spectral element code currently developed by P.\ Fischer (chief architect), J. Lottes, S.
Kerkemeier, A. Obabko, K. Heisey, O. Marin and E. Merzari at Argonne National Laboratory (ANL). 
Nek5000 has several strengths.  The code is fast (partly due to its efficient preconditioners) and
has run on over a million ranks on ANL's Mira supercomputer.  Because the code is based on spectral
elements, its numerical accuracy converges exponentially as the spectral order increases. Nek5000
also allows direct control over the parameters in this problem, including direct control of the
viscosity.

The simulation setup was as follows.  The simulations were in three dimensions with the flame
propagating in the $y$-direction against a gravitational force in the $-y$ direction. The domain was
a square shaft of the same length in the $x$- and $z$-directions and a much larger height in the
$y$-direction. The boundary conditions were periodic on the side walls. The top of the simulation
domain was subject to an inflow condition with $u_x=0$, $u_z=0$ and $u_y=-v_{\text{shift}}$, where
$v_{\text{shift}}$ was dynamically set to the flame speed calculated at each time step. This
procedure is permitted for this set of fluid equations by extended Galilean invariance
\citep{pope2000}. The changing inflow velocity held the flame surface at a fixed-on-average position
within the domain. The bottom of the domain was subject to an outflow condition in which a small
region at the bottom of the domain was made compressible so that all characteristics near the bottom
of the domain pointed out of the domain. We compared the results from this configuration with
simulations in which the bottom boundary was subject to an outflow condition with $u_x=0$, $u_z=0$
and $u_y=-v_{\text{shift}}$ and found that the bottom boundary condition did not make a substantial
difference to calculated average quantities like the flame speed. The temperature was held at $T=0$
(fuel) for the top boundary and $T=1$ (ash) for the bottom boundary. The flame surface remained
within the domain and did not approach either boundary.

The flame front for all of the simulations was a plane initially perturbed by a randomly-seeded
group of sinusoids with an amplitude of $3.0$ and wavenumbers between $k_{min}=4$ and $k_{max}=16$.
The initial temperature profile was given by $T(x,y,z) = 0.5( 1 - tanh(2 r(x,y,z)/\delta_t))$, where
$r(x,y,z) = y - q(x,z)$, where $q(x,z)$ is the position of the flame front including the effect of
the perturbation and $\delta_t$ is the initial width of the front which is $\delta_t = 4$ for the
bistable reaction.  The initial velocity was zero in the entire domain.

The parameters for all of the simulations are given in Table \ref{table-sims}.  In total there were
11 different combinations of parameters simulated:  six simulations in a domain of width $L=32$ with
$G=1,2,4,8,16,32$ and five simulations in a domain of width $L=64$ with $G=0.5,1,2,4,8$.  The
simulations varied in size depending on the resolution required to resolve the turbulent cascade and
to ensure that the velocity field downstream of the flame front would have adequate space for
evolution.  The total running time for each simulation was such that the flame speed would undergo
several oscillations of its dominant period after the flame reached a statistically steady state.
The flame speeds as a function of time in the statistically steady state are shown in Figures
\ref{fig-s-L32} and \ref{fig-s-L64}.  All averaged quantities were computed over the statistically
steady state and ignored the initial transient.

We confirmed that the simulations were resolved in several different ways.  First, we calculated the
expected viscous scale from the measured Reynolds number and ensured that the average resolution was
smaller than this value.  Second, we computed viscous scales in the three coordinate directions
directly from the velocity field gradients and ensured that the resolution was smaller than these
directional viscous scales.  In all cases, the viscous scale calculated directly from the measured
$Re$ was smaller than the smallest directional viscous scale (as expected).  Finally, we conducted
at least one resolution test for each simulation: a lower and a higher resolution test for the
smaller simulations, and a lower resolution test for the larger simulations.  In the worst case, the
difference between measured flame speeds for different resolutions was about six percent. Some of
the variability between simulations is due to the uncertainty associated with averaging over an
oscillating function (see Section \ref{ssec:measure}), but there also may be
an intrinsic variability due to slightly different realizations of the flame behavior with the same
parameter values.  In these ways, we have confirmed that the simulations are resolved and the
qualitative conclusions discussed in later sections do not depend on the resolution.

\section{Turbulent Flame Regimes and the Flame Width}
\label{sec:flameregimes}

In this section, we introduce the basic theory of the traditional turbulent combustion regimes and
show that the predictions of that theory do not match our results.  Traditional turbulent combustion
considers a flame consuming turbulent fuel; the behavior of the flame depends on how strong the
turbulence is. The physical mechanisms thought to underlie the turbulent combustion regimes also
form the basis of turbulent flame speed models (Turb-SGS). So, a test of whether these regimes apply
to Rayleigh-Taylor unstable flames is also an indirect test of the physical validity of Turb-SGS
models.  In addition, many models of the deflagration-to-detonation transition (DDT) rely on
Rayleigh-Taylor unstable flames transitioning from the flamelets regime to the reaction zones
regime. It is important to determine whether or not this transition actually occurs.

\begin{figure*}
\begin{center}
\includegraphics[height=6in,angle=270]{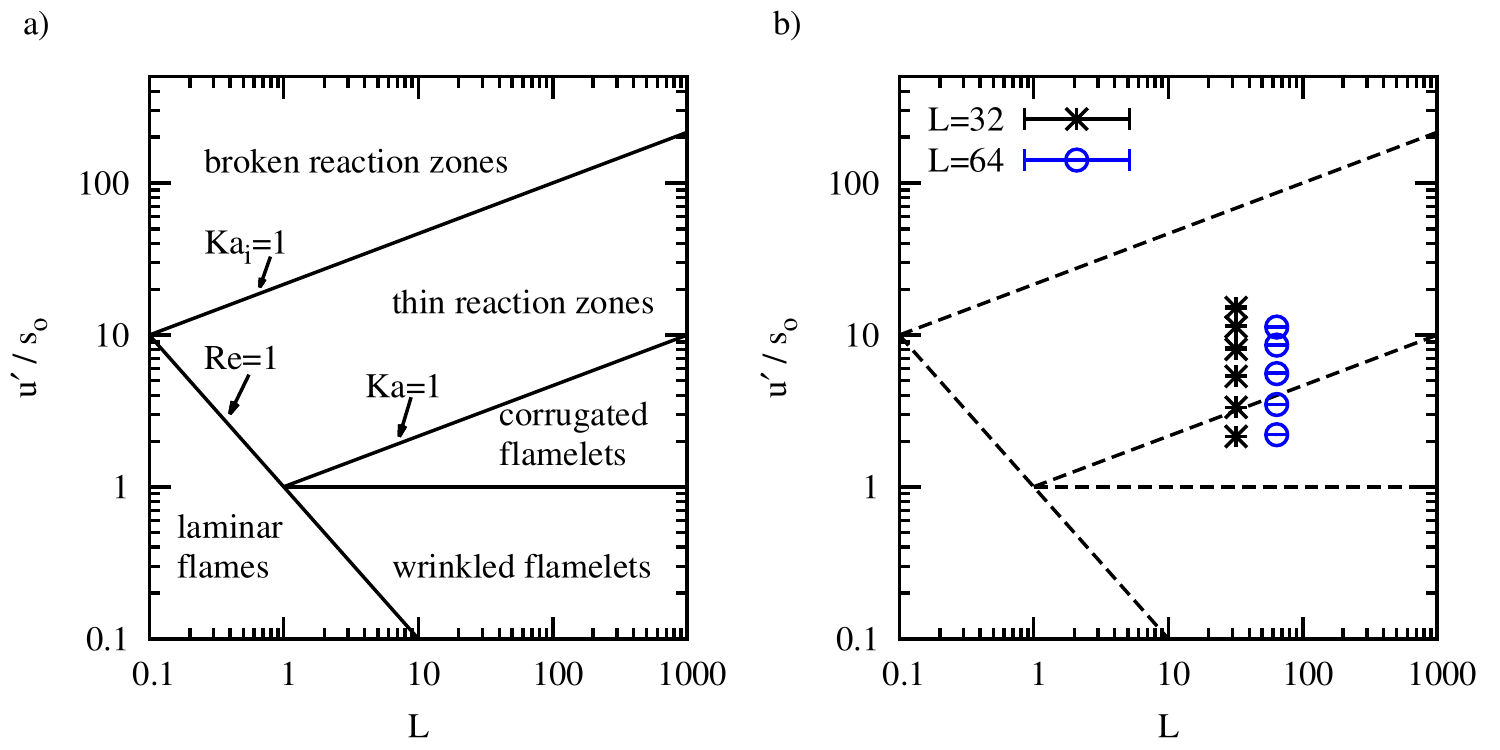}
\end{center}
\caption{Combustion Regimes Diagram. Part (a) shows a traditional turbulent combustion regime diagram
(adapted from \citet{peters2000}) with regimes based on comparisons between the time scales,
velocity scales and length scales of turbulence and a laminar flame.  Part (b) shows the positions of
the simulations on the regime diagram (blue circles and black asterisks) and the regime predictions
(as dotted lines). Most of the simulations are predicted to be in the thin reaction zones regime.}
\label{fig-regimes}
\end{figure*}

In turbulent combustion theory, it is common to define behavioral regimes based on velocity and length scale ratios.
Comparing various ratios leads to a regime diagram, illustrated in Figure \ref{fig-regimes}, part (a). There are five
different
major regions (this number can vary depending on the regime diagram): laminar flames, wrinkled flamelets, corrugated
flamelets, thin reaction zones and broken reaction zones \citep{peters2000}. When both $\ell / \delta$ and $u' / s_o$ are
small enough that $Re < 1$, the flame is laminar; it is not affected by turbulence and it remains flat with the laminar
temperature profile. For larger values of $\ell / \delta$, but $u' / s_o < 1$, the flame is in the wrinkled flamelets regime.
In this regime, the turbulent velocity is less than the laminar flame speed so the flame is practically unaffected by the
turbulence; the flame is close to laminar. If the turbulent velocity is larger than the laminar flame speed, turbulence will
affect the flame front. The details of the interaction depend on the ratio of the flame propagation time to the eddy turnover
time of the viscous scale eddies. This ratio is the Karlovitz number, $Ka = t_F / t_{\eta}$, which if $Pr = Sc =1$ also
compares the flame width to the size of the Kolmogorov scale or the velocity at the Kolmogorov scale to the laminar flame
speed: $Ka = (\delta / \eta)^2 = (v_\eta / s_o)^2$. If $Ka < 1$, the flame timescale is less than the Kolmogorov timescale
and the smallest viscous eddies are larger than the laminar flame width. In this regime, the corrugated flamelets regime, the
eddies wrinkle the flame front but do not change the basic internal laminar flame structure. On the other hand, if $Ka >1$
(the thin reactions zone regime), then the flame propagation time is longer than the Kolmogorov time and $\eta < \delta$ so
some eddies are smaller than the laminar flame width. In this case, it is thought that the eddies smaller than the laminar
flame width increase the local thermal diffusivity, thickening the flame.  Finally, the flame is in the
broken reactions zones regime if the turbulent eddies are smaller than the thin reaction zone within the flame; then, $Ka_i =
(\delta_i / \eta) ^2 > 1$.   In this regime, the flame may be entirely disrupted by turbulence and could extinguish.

\begin{figure*}
\begin{center}
\includegraphics[height=6in,angle=270]{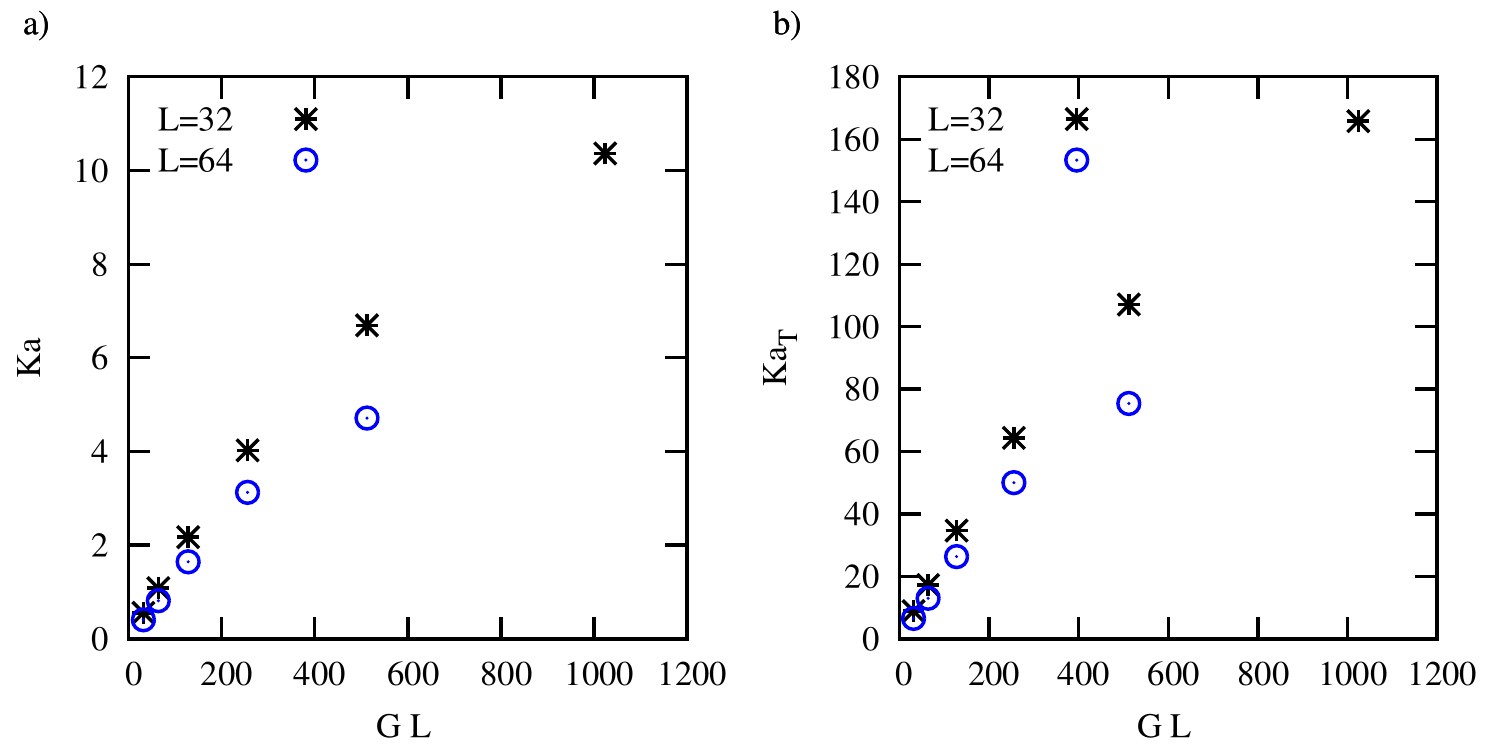}
\end{center}
\caption{Karlovitz Numbers.  Part (a): the traditional Karlovitz number,  $Ka = (\delta / \eta)^2$,
measured from the simulations.  Values over $1$ indicate that the simulation should be in the
reaction zones regime.  Part (b): the thermal Karlovitz number, $Ka_T = (4 \delta / \eta)^2$, measured
from the simulations. This is a Karlovitz number based on the laminar thermal width of the flame ($4
\delta$).  The large values of $Ka_T$ indicate that the viscous scale is much smaller than the
thermal flame width for many of the simulations. The Rayleigh-Taylor instability is stronger for
higher values of $GL$.}
\label{fig-Ka}
\end{figure*}

In order to check the validity of the regime diagram for RT unstable flames, it is first necessary to define the Karlovitz
numbers for bistable model flames.  These flames are thicker than more realistic model flames (e.g. the Arrhenius reaction)
for which the reaction vanishes exponentially at low temperatures. The bistable reaction has a much less extreme drop-off at
low temperatures, so the reaction is spread out over a larger physical space.  The innermost reaction zone, where the
reaction rate is fastest, is also larger for the bistable reaction than for the Arrhenius reaction.  Nevertheless, to
facilitate comparison with other simulations and experiments, we will use the standard definition of $Ka = (\delta / \eta)^2$
for most comparisons and define $Ka_i = (\delta_i / \eta) ^2$ (assuming that $\delta = 10 \delta_i$), although these choices
result in an underestimation of $Ka$ and $Ka_i$ for the bistable model flame.  To remedy this difficulty, we also define a
thermal Karlovitz number, based on the full thermal flame width $\delta_T = 4 \delta$ so $Ka_T = (4 \delta / \eta)^2$. 
$Ka_T$ is an indicator of whether turbulent eddies are able to penetrate the physical flame width.  Measurements of both $Ka$
and $Ka_T$ are shown in Figure \ref{fig-Ka}.  In addition, all of the simulations are shown on the regime diagram in Figure
\ref{fig-regimes}. 
 
According to the regime diagram and the measured values of $Ka$, a few of the simulated flames
should be in the corrugated flamelets regime, while most should be in the thin reaction zones
regime. Specifically, for $L=32$, flames with $G \lesssim 2$ should be flamelets and for $L=64$,
flames with $G \lesssim 1$ should be flamelets.  For all higher values of $G$, the flames are
expected to be in the reaction zones regime.  Traditional turbulent combustion theory predicts that
these flames should be thicker than the thermal laminar flame width (here, $\delta_T = 4$) because
eddies on scales smaller than the flame width should enhance thermal transport and thicken the
flame. 

\begin{figure*}
\begin{center}
\includegraphics[height=6in,angle=270]{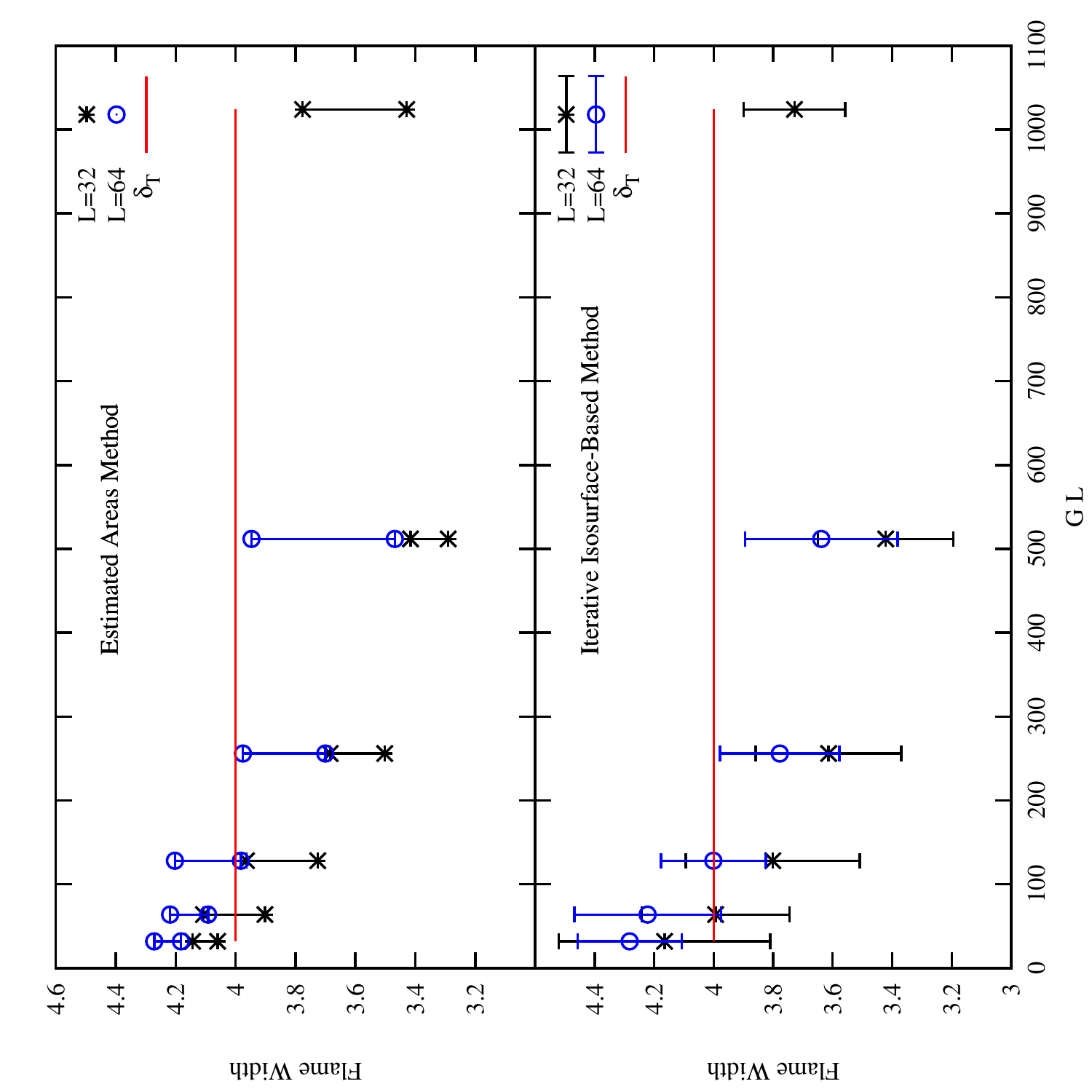}
\end{center}
\caption{Turbulent Thermal Flame Width vs. GL, calculated from the
simulations by two different methods. Top Panel: Estimated Areas Method. Bottom Panel: Iterative Isosurface-Based Method;
flame widths computed in post-processing. The laminar thermal flame
width is $\delta_T =4$ and is indicated by a solid red line. Surprisingly, most of the simulations
have a flame width smaller than the laminar thermal flame width, implying that although $Ka > 1$ the
flames are stretched flamelets instead of thin reaction zones.}
\label{fig-widths}
\end{figure*}

To check this prediction, we measured the flame width in two different ways.   Both methods involve
dividing the flame volume by an area to estimate the flame width.    In the first method, which we will call the ``estimated
areas method'',  we measured the total volume of material between the $T=0.1$ and $T=0.9$ temperature contours and then
divided that volume by two different indirect estimates of the flame surface area to find upper and lower estimates of the
flame width.   The first of these estimated areas is the flame surface area that would produce the measured turbulent flame
speed if the turbulent flame speed follows the relation $s/s_o = A / A_o$, where $A$ is the area of the turbulent flame and
$A_o$ is the area of the laminar flame.  This assumption may overestimate the flame area, as discussed in Section
\ref{ssec:cusps}, so this calculation gives a lower bound for the flame width.  An upper limit for the flame width is
calculated by assuming that the flame surface area is determined by the predicted Rayleigh-Taylor flame speed so $A =
\sqrt{1+0.125 GL}$.    We calculated both lower and upper bounds on the flame width at each time step and then calculated the
time-averaged bounds (excluding data from an initial transient period).   The flame width range measured using this method is
shown in the top panel of Figure \ref{fig-widths}.

The problem with dividing the entire flame volume by a representative
surface area is that this area must be correctly chosen.   In the estimated areas method, described above, we estimated
these areas indirectly using physical reasoning.   A more direct approach is to divide the isovolume by the surface area
of a representative temperature contour, for example, the $T=0.5$ contour, but this requires choosing
the ``correct'' contour.   Different temperature contours can have very different surface areas, adding to the
difficulty.

The second method, the ``iterative isosurface-based method'',  sidesteps these problems by using the surface areas
of temperature isosurfaces to estimate the flame width iteratively.   This method is described in detail and is
mathematically formulated by \citet{poludnenko2010}, see their Appendix A.     The iterative isosurface-based method
exploits the fact that isosurfaces with more similar $T$ values also have more similar surface areas.  For instance, the
$T=0.1$ isosurface area is much more similar to the $T=0.15$ isosurface area than to the $T=0.9$ isosurface area.    This
means that the flame width can be accurately estimated by dividing the total flame volume into smaller subvolumes bounded by
isosurfaces defined by similar $T$ contours.   Because these contours have similar surface areas, the average width of volume
that they bound can be estimated unambiguously.    Then, the average width of the entire flame is just the sum of the widths
of the smaller flame subvolumes.

Ideally, the flame would be divided into infinitely many subvolumes, but in practice the number of divisions is
limited by the resolution of the simulation.   Subvolume widths should be close to, but not substantially less than, the
resolution scale. This requirement suggests an algorithm in which the entire flame volume is divided into subvolumes, the
width of each subvolume is calculated and then, if any subvolume width is greater than some factor, $\alpha$, of the
resolution (\citet{poludnenko2010} used $\alpha=4$, we used $\alpha=2$), that subvolume is further divided iteratively. 
\citet{poludnenko2010} describe this algorithm in detail.    We changed their algorithm in one way;  \citet{poludnenko2010}
used the area of the isosurface on only one side of the each subvolume to calculate the subvolume width.   This is a
reasonable procedure if the resolution of the simulation is small enough that the isosurface areas of the bounding $T$
contours are nearly identical. However, it is easy to calculate upper and lower bounds for the width of any subvolume by
dividing the volume by the surface areas of both bounding isosurfaces.   These bounds for the widths of the subvolumes are
then added to get the total range of possible values for the total flame width.

We implemented the iterative isosurface-based method using the VisIt Python Interface  \citep{visit}.  VisIt uses
the marching cubes algorithm to construct contours and includes built-in queries for the isosurface areas and
isovolumes.   We ran our analysis code in post-processing, analyzing data files that were written out every tens to hundreds
of time steps during the original simulations.   Finally, we calculated a time-averaged flame width, using data from all the
files except for those corresponding to the initial transient.  The flame widths calculated using the iterative
isosurface-based method are shown in the bottom panel of Figure \ref{fig-widths}.

The two flame width calculation methods (see Figure \ref{fig-widths}) produced similar results for the time-averaged flame
width.  Surprisingly, the flame is thinner at larger values of $GL$ instead of thicker as predicted by turbulent combustion
theory. This implies that instead of being thickened by small-scale turbulent eddies, the flame is actually being thinned,
probably by the stretching action of the Rayleigh-Taylor instability. The flames have not entered the thin reaction zones
regime although $Ka > 1$ and $Ka_T >> 1$, instead they are stretched flamelets.

It is clear from these results that the traditional combustion regimes do not apply to RT unstable
flames for the parameter values studied.  Of course, it remains to been seen whether a transition to
reaction zones occurs at higher $Ka$.  It worth noting that even traditional turbulent flames
often do not show a transition at $Ka = 1$, although they may show a transition at higher $Ka$
\citep{driscoll2008}. This suggests that the theory is only approximate, even for traditional
turbulent flames.  However, thinning of a flame is highly unusual and suggests that the inner structure
of RT unstable flames is being determined by a straining mechanism (probably the RT instability) and
that the flames are not being affected internally by small eddies. This fits with the physical
picture of Rayleigh-Taylor unstable flames.  Vorticity is created by temperature gradients across
the flame front, and is not able to diffuse ahead of the flame (which has been confirmed by
measurements in these simulations).  If the vorticity is quickly driven downstream from the flame,
the flame front will not interact with smaller turbulent eddies at all.  Traditional turbulent
combustion is geometrically and physically different because the flame moves through
a turbulent fuel and is forced to interact with each individual turbulent eddy to propagate. 
RT flames do not have to interact with the turbulent eddies to propagate, so there is no reason to
expect that they generally behave like turbulent flames.  The observed thinning of the flame
suggests that traditional turbulent combustion regimes do not apply to RT unstable flames and that,
by proxy, flame speed models based on the physical ideas underlying the traditional turbulent
combustion regimes also may not apply.  We will directly compare some of these flame speed models to
our results in the next section.  Finally, these results imply that achieving DDT by transitioning
to the reaction zones regimes may not be possible, since the transition may not ever occur.

\section{The Flame Speed and Comparison With Flame Speed Models}
\label{sec:speedtests}

In this section, we will test the predictions of flame speed models using flame speed measurements
from the parameter study simulations.  Specifically, we will introduce and test several
turbulence-based models and the RT-based model.  Turbulence-based models generally give a
dependence of the turbulent flame speed on the rms velocity ($u'$) and other quantities. The models
we will test include linear, scale invariant, and power law models. For each model we will compare
the prediction of the model for the turbulent flame speed, $s$, in the entire domain with
measurements.  This procedure is not the most rigorous test of the subgrid models, which would
involve implementing the models in simulations with unresolved scales, but it is a good basic check.
Indeed, the scale at which the model is tested shouldn't matter because all types of subgrid models
currently in use are either compatible with the idea, or assume, that the flame surface is a
fractal.  The fractal nature of RT unstable flames was confirmed directly in 2D by
\citet{hicks2013}.

Surprisingly, there is not one, universally-used definition of the turbulent flame speed, making it
difficult to compare experimental and theoretical results in the field of turbulent combustion
\citep{lipatnikov2002,driscoll2008}.  It is even unclear whether there is one ``correct'' definition
of the turbulent flame speed; different definitions may be more useful in different circumstances.
In spite of these ambiguities, there is widespread agreement that the concept of a turbulent flame
speed is still a useful one.  There are at least two commonly used definitions of the global
turbulent flame speed \citep{driscoll2008}. The first definition, of the displacement speed,
measures the physical distance covered by a certain isosurface of the flame in a certain time.  This
definition is isosurface dependent, and the calculated flame speed can depend by a factor of 2-3 on
the isosurface chosen. The second definition, the global consumption speed, is based on the
measurement of the total amount of fuel consumed by the flame in a given amount of time and the area
of a chosen isosurface, so this measure is also isosurface dependent.  In this paper, we use a
third definition, the bulk burning rate \citep{vladimirova2003a}, which measures the global
production of reactants per unit time, but does not rely on measuring isosurface areas.  For our
simulations, the bulk burning rate is defined as
\begin{equation}
 s(t) = \frac{1}{L^2} \int_0^L \int_0^L \int_{-\infty}^\infty R(T) \,  dy dx dz .
\label{eqn:fspeed}
\end{equation}
The bulk burning rate is very similar to, but is less ambiguous than, the global consumption speed
and is preferred when measuring the flame speed for model flames for which $R(T)$ is known. For the
rest of this paper, we will refer to the bulk burning rate as the turbulent flame speed, $s$.

We will begin this section by giving a brief history and overview of models for the turbulent flame
speed in Section \ref{ssec:turb-hist} and then continue with a discussion of the specific turbulent
flame speed models that have been implemented and used in full-star Type Ia simulations in Section
\ref{ssec:turb-astro}.  After a short explanation of how the measurements were made (in Section 
\ref{ssec:measure}), we will compare the measurements of the turbulent flame speed to the
predictions of several turbulent flame speed models in Section \ref{ssec:compare}.
Finally, in Section \ref{ssec:rt} we will compare the turbulent flame speed measurements with
the predictions from the RT-based flame speed model.

\subsection{Turbulence Based Flame Speed Models: A History}
\label{ssec:turb-hist}

\citet{damkohler1940} was the first to make theoretical predictions of the turbulent flame speed and
assess those predictions with experiments.  By experimenting with Bunsen burner flames, he was able
to identify two basic regimes of turbulent combustion by comparing the diameter of the Bunsen burner
tube (which is the integral scale, $\ell$) and the width of the flame.  If $\ell > \delta$,
Damk\"{o}hler found that the turbulent flame speed could be fit by the relation $s = A \, Re + B$,
but if $\ell < \delta$ then $s \propto \sqrt{Re}$.  In the modern language of flame regimes,  $\ell
> \delta$ corresponds to the flamelets regime and $\ell < \delta$ to the thin reaction zones regime.
 Theoretically, Damk\"{o}hler considered the physical cause of these scaling laws.  He reasoned that
turbulence increased the surface area of the flame and, therefore, the flame speed, so that 
\begin{equation}
 \frac{s}{s_o} = \frac{A}{A_o} .
\end{equation}

Considering the geometry of the Bunsen burner flame, Damk\"{o}hler suggested that the wrinkling of
the flame surface is proportional to $u'$, so that for large values of $u'$, $s/s_o\propto u'$. 
Taking into account the requirement that the turbulent flame propagates at the laminar flame speed
if $u'=0$ this becomes
\begin{equation}
 \frac{s}{s_o} = 1 + C u'
 \label{eqn:Dam}
\end{equation}
where C is a constant.  This expression for the flame speed is still used today and fits many
observations surprisingly well.   

When $\ell < \delta$, in the thin reaction zones regime, Damk\"{o}hler hypothesized that the
smallest scale turbulence would not wrinkle the flame, but instead would enhance small-scale
microscopic transport within the burning region.  Then the turbulent flame speed is $s \approx
(\kappa_T \alpha)^{1/2}$ where $\kappa_T$ is, in modern terms, the turbulent diffusivity and
$\alpha$ is the reaction rate.  This is the same as the equation for the laminar flame speed, $s
\approx (\kappa \alpha)^{1/2}$, but $\kappa$ is replaced by $\kappa_T$.  Then $s / s_o = (\kappa_T /
\kappa) ^{1/2}$ and finally
\begin{equation}
   \frac{s}{s_o} \sim \left( \dfrac{u' \ell}{s_o \delta} \right)^{1/2},
\end{equation}
which reduces to $s / s_o \propto \sqrt{Re}$ if $Pr = 1$.  Summarizing, Damk\"{o}hler's predictions
for
the dependence of the turbulent flame speed on $u'$ are $s \propto u'$ if $\ell > \delta$ and $s
\propto \sqrt{u'}$ if $\ell < \delta$.

Since Damk\"{o}hler's time, there have a been many thousands of experimental measurements of the
turbulent flame speed and tens of theoretical expressions formulated to explain the experimental
results. A full history is beyond the scope of this paper; more information can be found in the many
articles and textbooks that review the subject, for example see
\citet{bray1980,bray1990,bradley1992,peters2000,lipatnikov2002,bilger2005,law2006,driscoll2008,
kuo2012,lipatnikov2013,poinsot2013}. We will only cover a few of the key points.

\begin{figure}
\begin{center}
\includegraphics[height=2.5in,angle=0]{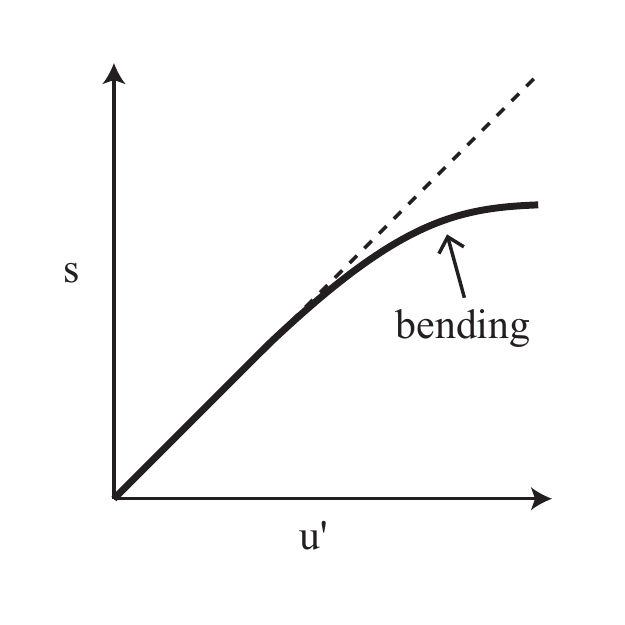}
\end{center}
\caption{The Bending Phenomena. This cartoon shows ``bending'' in the dependence of the turbulent
flame speed, $s$, on the turbulent rms velocity, $u'$.  Bending refers to a deviation from
linear dependence that results in a concave-down curve.  This choice of axes is often referred to as
a ``burning velocity diagram''. }
\label{fig-bending-cartoon}
\end{figure}

As experimental measurements of the flame speed were accumulated, a few basic properties of
turbulent flames were noticed. First, the basic scaling $s \propto u'$ generally fit the data well
at low values of $u'$, but at higher values of $u'$ the linear dependence failed.  The overall shape
change from a straight line on a $s$ vs. $u'$ plot (known as a burning velocity diagram) to a
concave down curve at high values of $u'$ is known as the ``bending phenomena'' (illustrated in
Figure
\ref{fig-bending-cartoon}).  It has been suggested that the bending behavior begins when the flame
enters the thin reaction zones regime.  Another clear experimental result is that there is not one
``turbulent velocity scaling law'' with fixed constants that fits all experiments. In fact, data
points are widely scattered on the burning velocity diagram, with some experiments showing lots of
bending and some showing none at all.  Nevertheless, attempts were made to find a best average model
by fitting the data aggregated from many experiments.  Researchers have also compiled
list of qualitative trends held in common by all or most experiments. After an analysis of a large
number of experimental databases, \citet{lipatnikov2002} identified several basic trends, the first
of which is an increase of $s$ with $u'$ as $s \sim {u'}^q$ with $q \approx 0.5$ being most likely
(i.e. bending occurs for most flames).  The other two basic trends are that $s$ and $ds /du'$ are
increased by $s_o$ and by pressure.  Overall, it is clear that predicting the exact turbulent flame
speed for an unknown system from first principles is exceedingly difficult.

Just as there have been many experimental measurements of the flame speed, there have been many
theoretical models formulated to explain these experimental measurements.  In general, many of these
models are based on a set of physical assumptions about the way that turbulence should interact with
flames, which, once combined, give an expression for $s$ in terms of $u'$ and, often, other system
parameters such as the integral scale and the laminar flame speed.  Some examples of approaches
include assessing the kinematic effect of turbulence on wrinkling directly
\citep{damkohler1940,shchelkin1943,clavin1979,peters1999}, modeling the flame as a fractal
\citep{gouldin1987,kerstein1988}, requiring the model to preserve scale invariance
\citep{pocheau1992,pocheau1994}, considering random exchanges of state between burned and unburned
cells (the pair-exchange model of \citet{kerstein1986}), and modeling the interaction between
turbulence and the flame as a series of vortex-flame interactions
\citep{meneveau1991,duclos1993}. All of these approaches have been shown to produce acceptable fits
to experimental data in some (but not all) cases.  In certain regimes various models can be shown to
be equivalent \citep{bray1990}.  A list of many of these models is given in \citet{lipatnikov2002},
Appendix B and in \citet{kuo2012}, Table 5.1. \citet{lipatnikov2002} also compare various models to
the experimental trends and highlight how well the models succeed (see their Table 2).

Currently, it is thought that a single, universal scaling law for $s(u')$ that applies to all
premixed turbulent flames across different combustion regimes probably does not exist.  Factors like
flame stretch, apparatus geometry, flame instabilities (such as the Landau-Darrieus instability),
quenching and the details of the reaction chain may influence the turbulent flame speed. Because of
these factors, a new focus has been to break premixed flames into categories by geometry and work to
understand each category separately.   Four common categories are envelope flames, oblique flames,
flat flames and spherical flames \citep{driscoll2008}. Unfortunately, none of these
categories is a good fit for the geometry of Type Ia flames, so it is necessary to study Type
Ia flames as a new category of premixed combustion.  Direct study is necessary because the theory of
premixed turbulent flames is not well-developed enough to predict scaling laws for new geometries
from first principles. This is partially because it has not been possible to determine the ``correct
physics'' by simply finding a working theoretical model since many models fit the data to some
extent. A final complication is that the the bending phenomena is still not well-understood. Bending
could be caused by a transition from flamelets to distributed burning, the merging and extinguishing
of flamelets due to strain, gas expansion or geometrical effects \citep{driscoll2008}. It is also
possible that different causes of bending are in effect for different geometries.  Overall, it is
now clear that a single flame speed model, $s(u')$, that is valid for all premixed turbulent flames
is unlikely to exist.
\clearpage

\subsection{Astrophysical Flame Speed Models: History and Discussion}
\label{ssec:turb-astro}

While it is clear that full-star Type Ia supernova simulations must incorporate a subgrid model, it
is far from clear what that subgrid model should be.  There have been three main attempts to adapt
turbulent combustion theory for the Type Ia problem: a simple linear scaling law ($s=u'$), a more
complex LES treatment including a flame speed scaling law derived from considerations of scale
invariance, and, finally, a flame speed scaling law meant to reproduce the bending seen in
terrestrial flames.   In this section, we will briefly discuss these models and distill their basic
characteristics, which we will test against our simulations in Section \ref{ssec:compare}. We
will save a discussion of the RT-based subgrid model until Section \ref{ssec:rt}.  

In general, the choice of a subgrid model for Type Ia simulations is very difficult.  First,
geometrically there is no simple terrestrial analog from which a turbulent flame speed model might
be taken.  Consequently, there have been no terrestrial laboratory experiments which directly
test ideas about Type Ia flames. Second, if the turbulent flame speed does depend on geometrical
history (see \citet{driscoll2008}) then determining the turbulent flame speed from first principles
could be almost impossible because conditions in the white dwarf and laminar flame properties change
during the explosion process. Third, Type Ia flames are highly unstable to the Rayleigh-Taylor
instability but the terrestrial flames from which the intuition of turbulent combustion has been
developed are mostly either hydrodynamically stable, unstable to the Landau-Darrieus instability or
unstable to various thermo-diffusive instabilities.   Finally, the turbulence generated by the RT
instability is downstream of the flame front and therefore probably affects the flame front
differently than turbulence that is initially upstream of the flame front.  In addition, it is
likely that the turbulence generated by the RT instability is not homogeneous and isotropic,
especially on large scales.  All of these factors make adapting models from traditional turbulent
combustion -- which deals with a stable flame propagating through uniform turbulence -- especially
difficult.

\citet{niemeyer1995} first incorporated a turbulence-based subgrid model into Type Ia supernovae
simulations by assuming a form for the turbulent flame speed at a given subgrid scale $\Delta$ of
\begin{equation} 
s = s_o \left( \frac{u(\Delta)}{s_o} \right)^ n 
\end{equation} 
with $1/2 \leq n\leq 1$ treated as a free parameter based on the value of the Gibson scale.  For
$n=1$, this result is equivalent to the classical Damk\"{o}hler relation $s \approx u'$ applied at
the scale $\Delta$. Next, \citet{schmidt2005, schmidt2006a, schmidt2006b} implemented a full LES
(Large Eddy Simulation) model of turbulent energy evolution on unresolved scales based on a Germano
decomposition filtering approach with localized eddy-viscosity and gradient-diffusion closures.  In
this approach, the turbulent flame speed at a given, unresolved scale $\Delta$ is derived from the
subgrid-scale turbulent energy $k(\Delta) = (1/2) {u'(\Delta)}^2$ and the flame speed relation
\begin{equation}
\label{eqn:SI}
 \frac{s}{s_o} = \left[1 + C_t \left( \frac{u'(\Delta)}{s_o} \right)^n \right]^{1/n}
\end{equation}
derived by \citet{pocheau1992,pocheau1994}.  Here $n=2$ was chosen to enforce energy conservation,
although this assumes that the turbulence has a Gaussian PDF, which is untrue for many turbulent
flows \citep{pocheau1994}.  $C_t$ is a tunable parameter which was set equal to either $1$ or $4/3$
(see also \citet{peters1999}). This form of the turbulent flame speed equation is scale invariant in
functional space which means that the interaction between the front geometry and the turbulent flow
is scale invariant.  Although scale invariance is a basic, if unstated, assumption of many
turbulence-flame interaction models, most models violate this property.  The scale invariant model
is an inexact choice for Type Ia supernova flames because they are RT unstable and Equation
\eqref{eqn:SI} was derived for hydrodynamically stable flames in homogeneous and isotropic
turbulence. \citet{pocheau1992,pocheau1994} explicitly warns that the property of scale invariance
may not hold for unstable flames.  In terms of limiting behavior, Equation \eqref{eqn:SI} reduces to
$s \approx \sqrt{C_t} u'$ when turbulence is strong, so this model does not produce the bending
phenomenon, which is fundamentally not scale invariant.  Numerically, the flame front is propagated
by a level set method using the G-equation. The Schmidt model explicitly includes an approximation
for the small-scale generation of buoyancy.

\begin{figure*}
\begin{center}
\includegraphics[height=6.3in,angle=270]{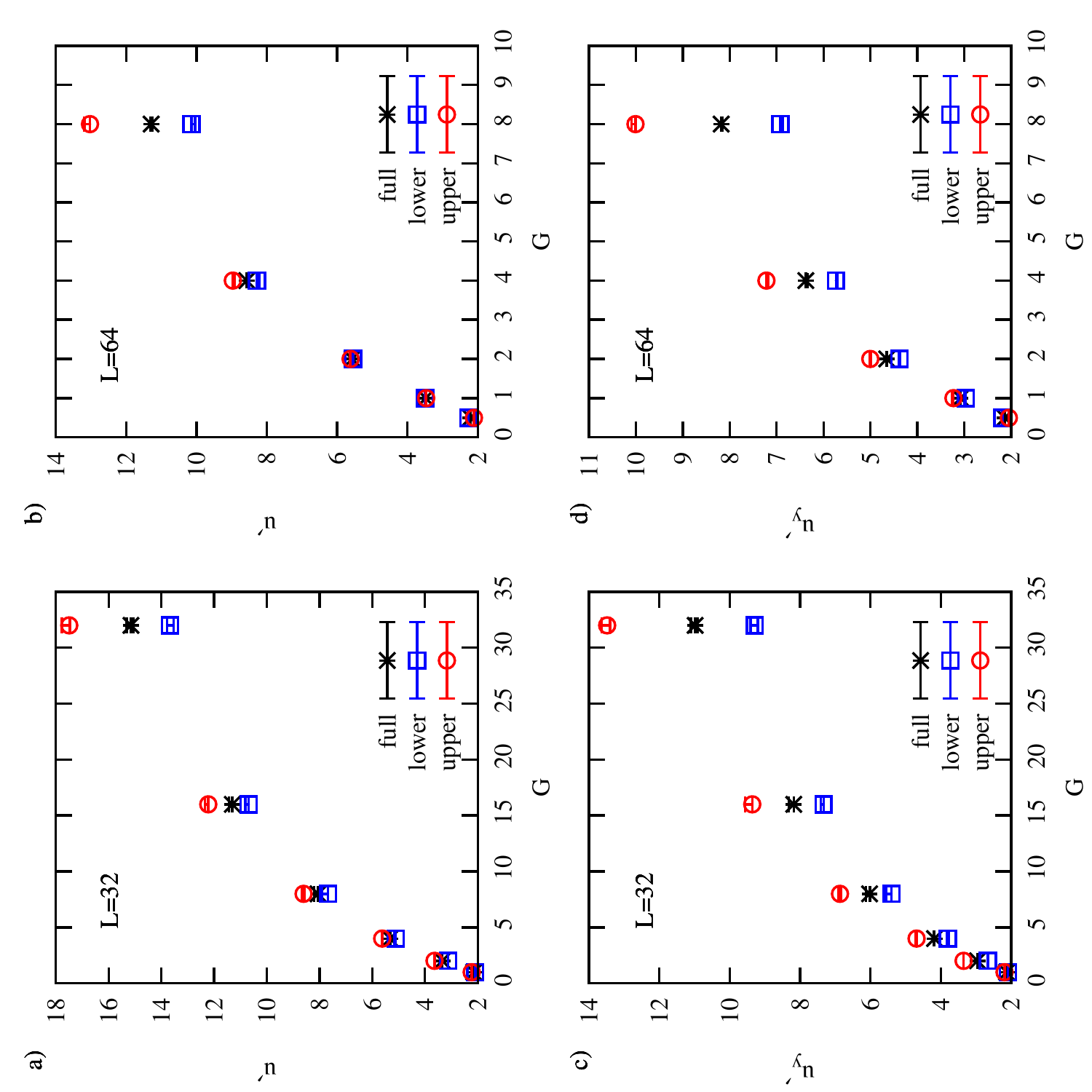}
\end{center}
\caption{ Turbulent Velocities vs. $G$ for $L=32$ and $L=64$.   These plots
show $u'$ and $u_y'$ defined by averaging over three different spatial regions: the full flame brush (black asterisks), the
top half of the flame brush (red circles), and the bottom half of the flame brush (blue squares).   Of these three
definitions, $u'$ (and $u_y'$) averaged over the full flame brush is the only truly global measure of the turbulent velocity.
 The other two measurements are shown as a general indication of the variation of $u'$ with height.   Generally, $u'$ is
larger in the upper half of the flame brush and the difference between the upper half and lower half $u'$ measurements is
larger at higher $G$.  Averages were computed in post-processing.}
\label{fig-urms-diff}
\end{figure*}

\begin{figure*}
\begin{center}
\includegraphics[height=5.3in,angle=270]{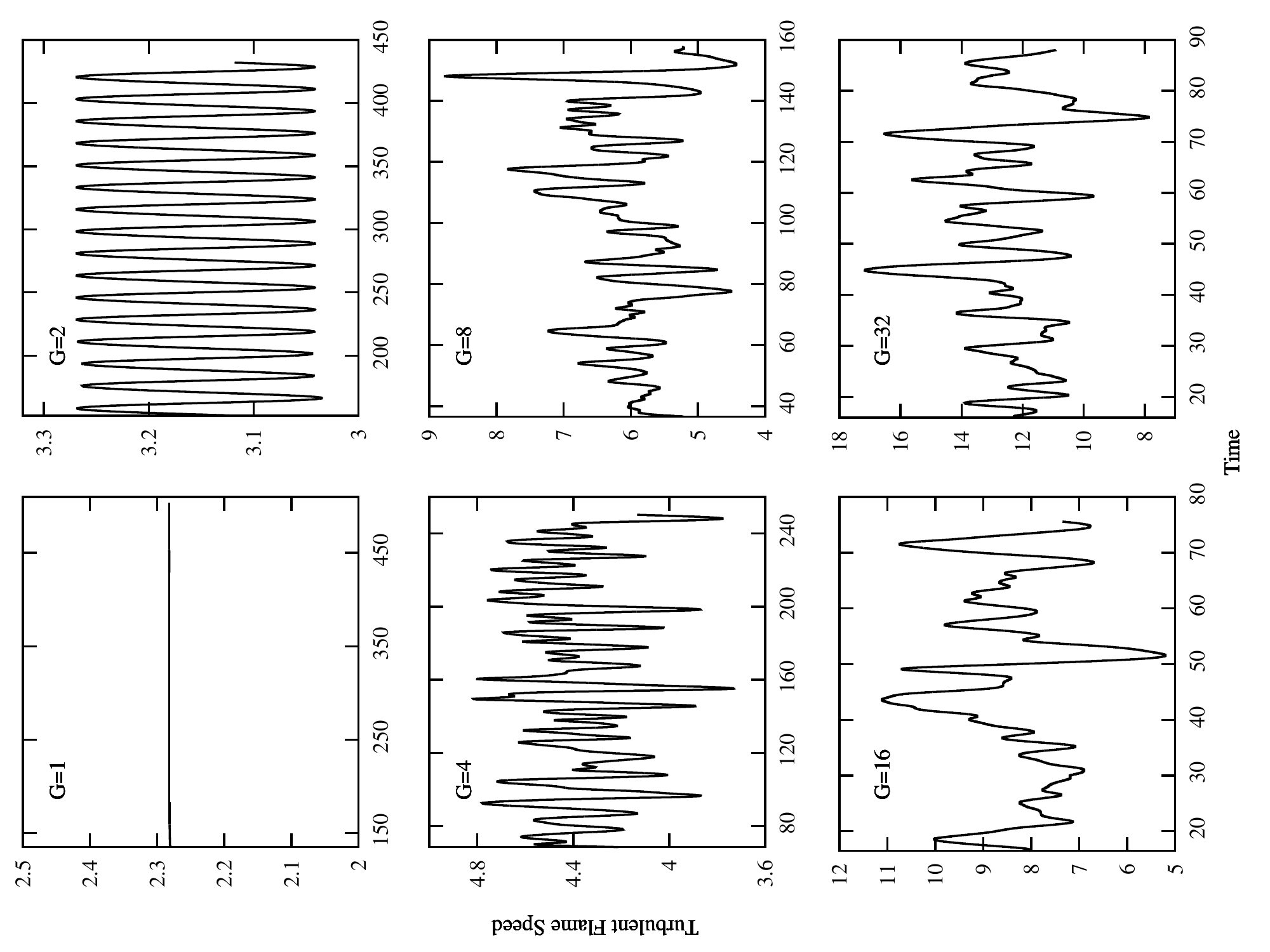}
\end{center}
\caption{Flame Speeds, $L=32$.  The turbulent flame speed, $s(t)$, as a function of time for the
simulations with a domain width of $L=32$.  The initial transient behavior is omitted from the plots
and only the saturated oscillation of the flame speed around a stationary mean is shown.  We
calculate the average $s$ for each simulation from the data in these time intervals.  Generally, the
flame speed is larger and oscillates more wildly for higher values of
$G$ at a fixed $L$. }
\label{fig-s-L32}
\end{figure*}

\begin{figure*}
\begin{center}
\includegraphics[height=5.3in,angle=270]{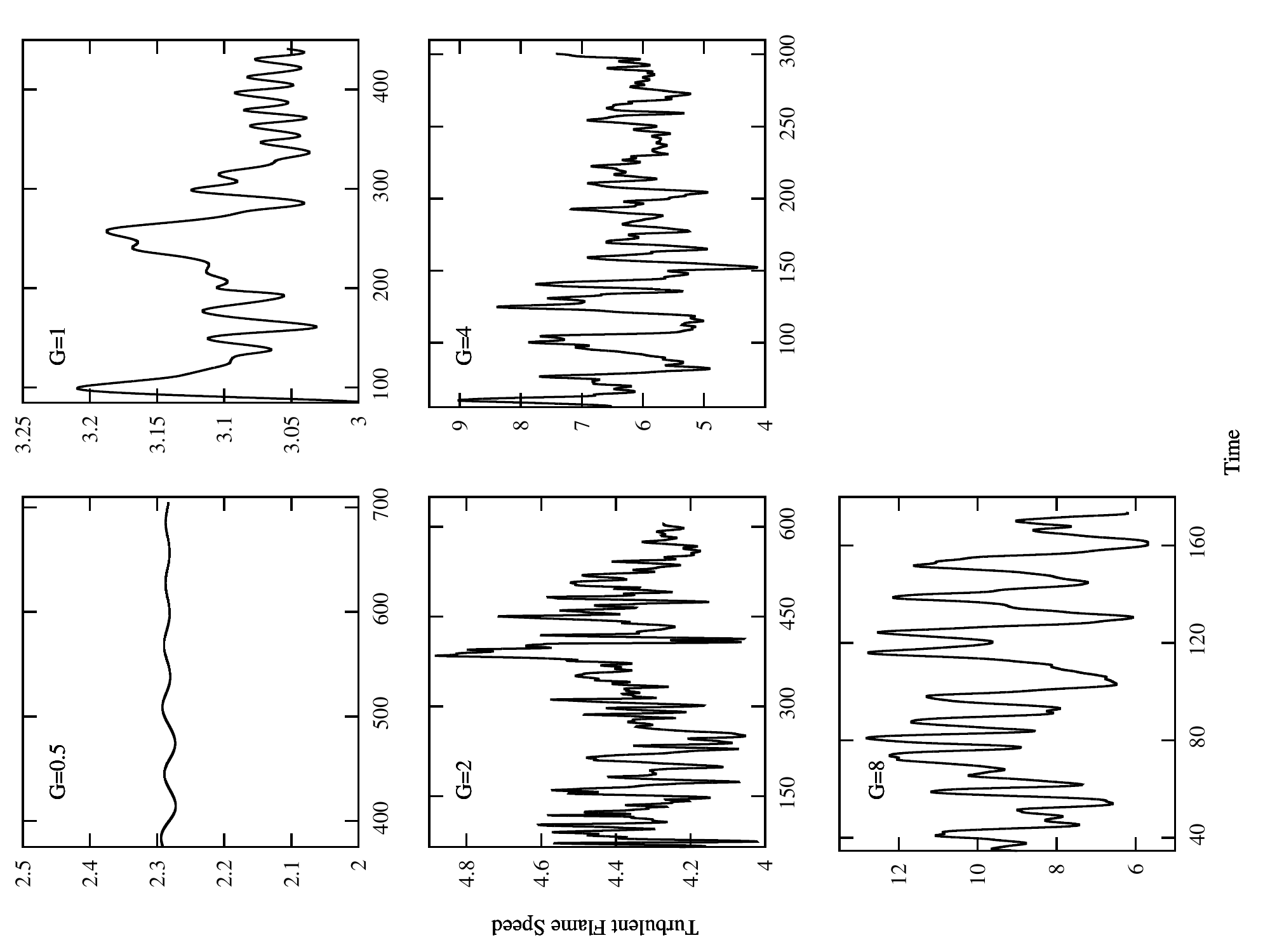}
\end{center}
\caption{Flame Speeds, $L=64$.  The turbulent flame speed, $s(t)$, as a function of time for the
simulations with a domain width of $L=64$.  The initial transient behavior is omitted from the plots
and only the saturated oscillation of the flame speed around a stationary mean is shown.  We
calculate the average $s$ for each simulation from the data in these time intervals.}
\label{fig-s-L64}
\end{figure*}

\begin{figure*}
\begin{center}
\includegraphics[height=6in,angle=270]{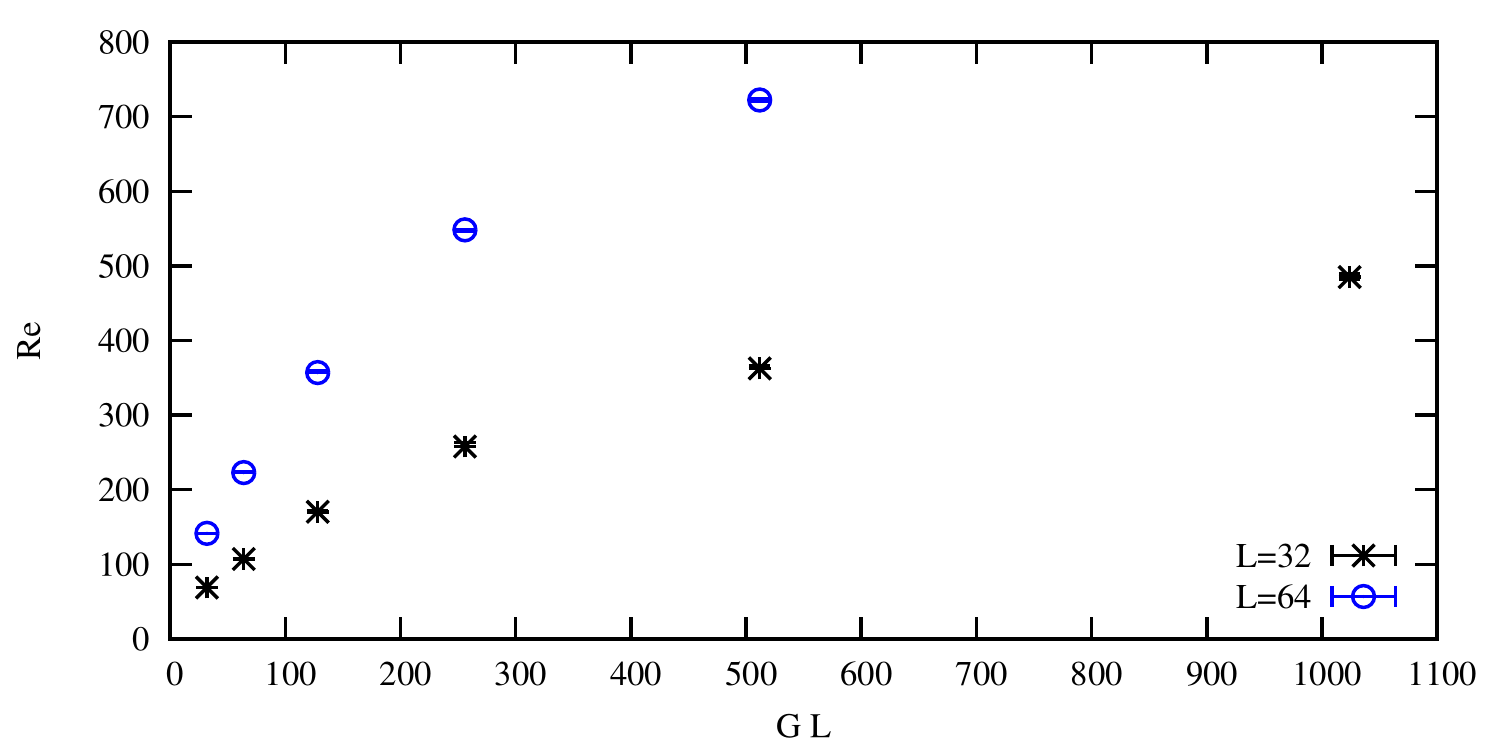}
\end{center}
\caption{Reynolds Number versus $GL$.  The figure shows the Reynolds number based on the
time-averaged $u'$ for each simulation, so $Re = u' L$ for $Pr =1$.  The black asterisks represent
the simulations with a domain width of $L=32$ and the blue circles represent simulations with a
domain width of $L=64$. Each simulation is represented by one point on the plot.  The product $GL$
is a basic measure of the strength of the Rayleigh-Taylor instability; larger $GL$ implies a
stronger RT instability. The simulations range from laminar to moderately turbulent.}
\label{fig-Re}
\end{figure*}

In contrast to the scale-invariant model implemented by \citet{schmidt2005, schmidt2006a,
schmidt2006b},  \citet{jackson2014} recently formulated an LES turbulence-flame interaction model
specifically to account for bending at high turbulence levels.  This model adapts an LES
model for terrestrial flames \citep{colin2000,charlette2002, charlette2002a} 
\begin{equation}
 \frac{s}{s_o} = \left ( 1 + \frac{\Delta}{\eta_c} \right) ^ \beta
\end{equation}
where $\eta_c$ is a cutoff scale for the wrinkling process and $\beta$ is a wrinkling exponent that
could depend on scale.  This model reduces to other models in various limits and can produce either
the Damk\"{o}hler scaling ($s \approx u'$) or bending, depending on $\beta$ and on $\eta_c$.
Physically,
$\eta_c$ is the inverse mean curvature of the flame which depends on an efficiency function $\Gamma$
which, in turn, depends on the net straining of all scales below $\Delta$.  This dependence is
found by assuming a balance between flame surface creation due to wrinkling and destruction due to
flame propagation and diffusion. In the model, $\Gamma$ is parameterized using the vortex-flame
interaction measurements of \citet{meneveau1991}. In other words, all interactions below the grid
scale $\Delta$ are assumed to be equivalent to the summed action of vortex-flame interactions on all
of those scales.  This model can take quenching into account by enforcing the requirement that
$\eta_c > \delta$.  Numerically, flame propagation is achieved by the propagation of a thickened
flame model. The current version of the model assumes that the turbulence is homogeneous and
isotropic and does not take into account the effects of the Rayleigh-Taylor instability.

Although the two main LES models currently in use
\citep{schmidt2005,schmidt2006a,schmidt2006b,jackson2014} are relatively complex, they make fairly
straightforward assumptions about the effects of turbulence on the flame.  Both models assume that
turbulence-flame interaction can be quantified using models adapted from terrestrial flame theory.
These models either mostly or entirely ignore the effects of the Rayleigh-Taylor instability. The
formulations of both models assume that the non-homogeneous and non-isotropic RT-generated
turbulence downstream of the flame front interacts with the flame front in the same way as
homogeneous and isotropic turbulence initially upstream of the flame surface would.  In the next
section, we will test the basic assumptions about the turbulent flame speed used in these models.
Both models predict that the flame speed will either scale roughly as $s \approx u'$ or as $s
\approx {u'}^n$ with $n < 1$ to reproduce the bending behavior seen in terrestrial flame
experiments.

\subsection{Flame Speed Measurements}
\label{ssec:measure}

In order to assess the various models for the turbulent flame speed, we measured both the turbulent
flame speed, $s(t)$, using Equation \eqref{eqn:fspeed} and the root-mean-square (rms) turbulent
velocity, $u'(t)$, for each simulation in the parameter study.  The turbulent flame speed as a
function of time for the $L=32$ and $L=64$ simulations is shown in Figures \ref{fig-s-L32} and
\ref{fig-s-L64} respectively. The initial transient growth of the flame speed as the instability
first develops is not shown. A few basic trends are apparent from the figures.  First, the flame
speed increases for larger $L$ or $G$ because flames with higher $GL$ are more unstable to the
Rayleigh-Taylor instability and also generate more turbulence.  Second, the size of the flame speed
oscillations grows as $GL$ increases because the flame goes through more severe cycles of flame
surface creation and destruction.  For example, the flame speed for $L=32$, $G=1$ is constant
because the flame surface is just a stable rising bubble, but the flame speed for $L=32$, $G=32$ is
very oscillatory and complex because the flame is strongly deformed by the RT instability (see
Figure \ref{fig-L32-temps}).  

For each simulation, we computed the average value of the flame speed, $s$, after excluding the initial transient, so that
each point in $L$,$G$ parameter space is associated with one averaged value of $s$.  It is this time-averaged value of $s$
that we compare to model predictions in the next section.  To estimate the uncertainty associated with the averaging process,
we calculated a running average error using the following procedure.  For every point in the time series (excluding the
initial transient), we computed the flame speed average using that point and all previous values of $s(t)$, so that as more
data was added to the time series the computed averaged $s$ changed less. We considered the averaging ``error'' to be the
range of averaged $s$-values computed as the last quarter of the time series points were added to the averaged data.  This
range of values is the uncertainty associated with averaging over a finite interval of an oscillating time series. In
general, these errors are relatively small, which indicates that the times series are long enough to calculate a meaningful
average.  These error bars are shown in each plot.   

 In general, time-averaged quantities were calculated using the data written out at every time step
during the simulation.   However, some averages were calculated later, after the simulation was complete, using time
snapshot data files which were written out on intervals of tens to hundreds of time steps.   Averages computed using the data
files alone are very close to averages computed at every time step.   These averages are indicated as
``computed in post-processing'' in the relevant figures to distinguish them from averages computed at every time step.

An averaged turbulent rms velocity, $u'$, was also computed for each simulation.  In order to get a
representative value of $u'(t)$ at a given time, we used the formula
\begin{equation}
 u'(t) = \sqrt{ < {u_x(t)}^2 + {u_y(t)}^2 + {u_z(t)}^2 > }
\end{equation}
where $< >$ indicates the spatial average over the volume between the top-most and bottom-most
extent of the $T=0.5$ to $T=0.8$ contour range that also satisfies the criterion $T > 0.5$.  In
other words, $u'(t)$ is based on spatial averaging in the ashes.

Our definition of $u'$ is clearly somewhat arbitrary: first, it depends on the selection of the
temperature range $T=[0.5,0.8]$.   However, measured values of $u'$ do not depend strongly on the temperature interval
selected;  we repeated the calculation of $u'$ using the interval $T=[0.1,0.9]$ and found similar results.   Second, our
definition of $u'$ is based on averaging over the whole flame brush, which does not explore the variation of $u'$
with height.    This is in keeping with the goal of this section: to compare the global flame speed to flame speed models
based on a global measurement of the turbulent velocity.  But, although our global definition of $u'$ is adequate for this
purpose, the overall vertical variation of $u'$ with height is still of great interest because RT unstable flames are
expected to have much more vertical variation than turbulent flames.    To give an idea of this variation, we show three
different measurements of $u'$  in parts (a) and (b) of Figure \ref{fig-urms-diff}.    The points represented by black
asterisks
show our measurements of $u'$ averaged over the entire flame brush (this is the data that will be used for model comparisons
throughout the rest of the section).   Red circles show $u'$ averaged only over the top half of the flame brush, and blue
squares show $u'$ averaged only over the bottom half of the flame brush.  In general, $u'$ is larger in the upper part of the
flame brush than in the lower part.  This difference is small at low $G$ and becomes significant at high $G$.   In general,
the variation of $u'$ with height does not affect any of the qualitative conclusions in this paper.    We plan to explore
vertical profiles of flame data more thoroughly in a future paper.

We also calculated directional rms velocities using 
\begin{subequations}
\begin{equation}
  u_x'(t) = \sqrt{ < {u_x(t)}^2> }
\end{equation}
\begin{equation}
  u_y'(t) = \sqrt{ < {u_y(t)}^2> }
\end{equation}
\begin{equation}
  u_z'(t) = \sqrt{ < {u_z(t)}^2> }
\end{equation}
\end{subequations} 
Time-averaged values, $u', u_x', u_y', u_z'$ were calculated using the same time-averaging procedure and averaging error
method as for the calculation of $s$.  Measurements of $u_y'$ are shown  in parts (c) and (d) of Figure
\ref{fig-urms-diff}.   The variation of $u_y'$ with height does not affect the qualitative conclusions in this paper, except
where noted. The Reynolds number for each simulation was calculated using $Re = u' L / Pr$, in dimensionless variables. The
Reynolds numbers for the simulations ranged between $70$ and $720$ showing that the simulations ranged from laminar to
moderately turbulent, see Figure \ref{fig-Re}.

\subsection{Turbulence-Based Flame Speed Models Comparisons}
\label{ssec:compare}

In this section, we will compare our measurements of the time-average flame speed ($s$) and the
time-averaged rms velocity ($u'$) to various models.  We consider three different basic types of
models: simple linear models, scale invariant models and models that reproduce bending.  As
described previously, each of these classes of model represents a basic type of model used in
traditional turbulent combustion theory and as subgrid models in astrophysical Type Ia simulations.

\begin{figure}
\begin{center}
\includegraphics[height=3in,angle=270]{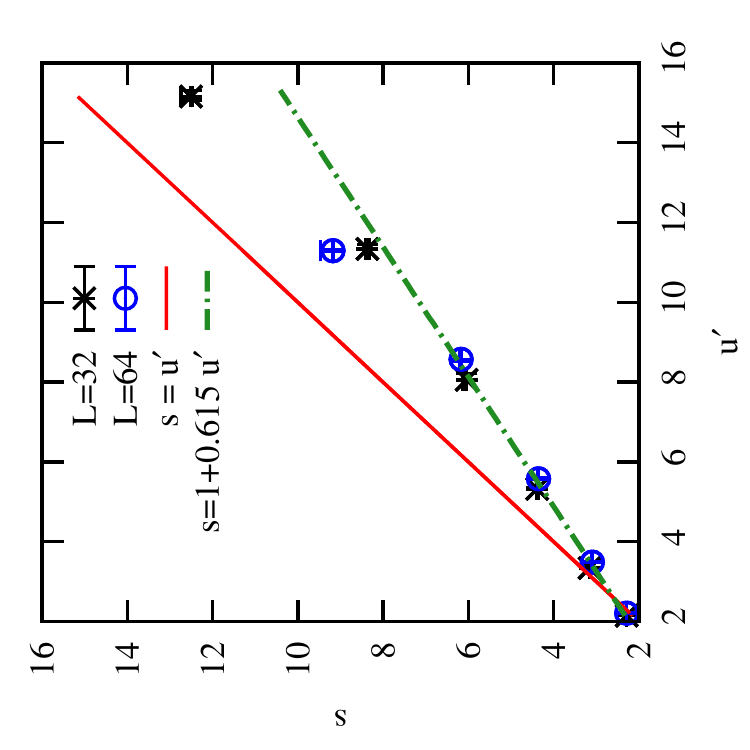}
\end{center}
\caption{Burning Velocity Diagram: Linear Models. The time-averaged turbulent flame speed ($s$) as a
function of the time-averaged rms velocity ($u'$) in the region downstream of the flame front. Data
from each of the $L=32$ and $L=64$ simulations are shown as black asterisks and blue circles
respectively.  Error bars are based on the cumulative time averaging procedure discussed in Section
\ref{ssec:measure} and represent the uncertainty associated with averaging over a finite oscillatory
time series.  Also shown are two simple linear model predictions. $s=u'$ is shown as a solid red
line and $ s= 1 + C u'$ with a least-squares fit for the value of $C$, $C=0.615$ is shown as a dashed
green line. Neither model is a good fit for the all of the data points. }
\label{fig-linearmodels}
\end{figure}

\begin{figure}
\begin{center}
\includegraphics[height=3in,angle=270]{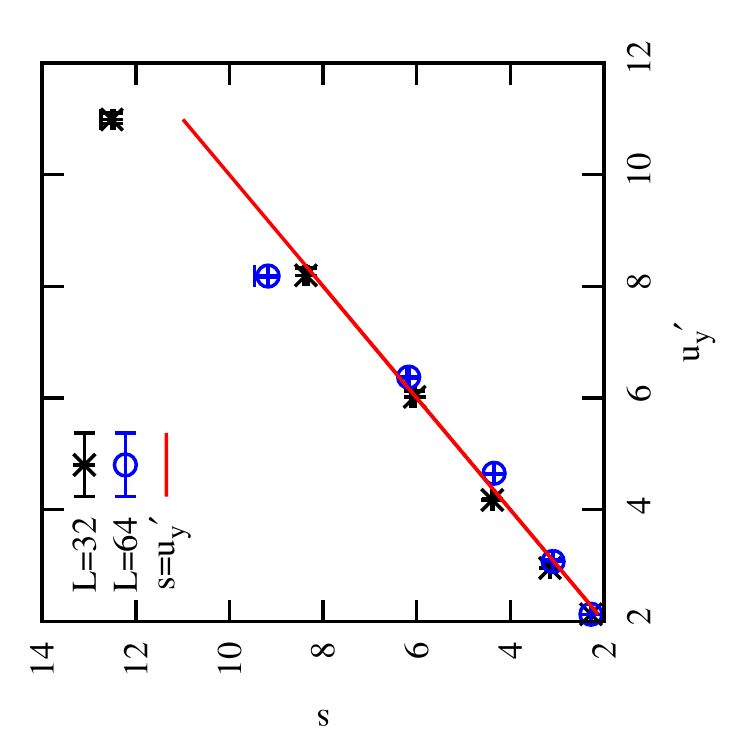}
\end{center}
\caption{Burning Velocity Diagram based on $u_y'$.  This figure shows a comparison between the measured time-averaged flame
speed ($s$) and the rms velocity in the direction of flame propagation ($u_y'$).  $s=u_y'$  (solid red line) fits the
simulation measurements well for small values of $u_y'$ (with no fitting parameter), but underestimates the flame speed for
large values of $u_y'$.}
\label{fig-urmsy-model}
\end{figure}

\begin{figure}
\begin{center}
\includegraphics[height=3in,angle=270]{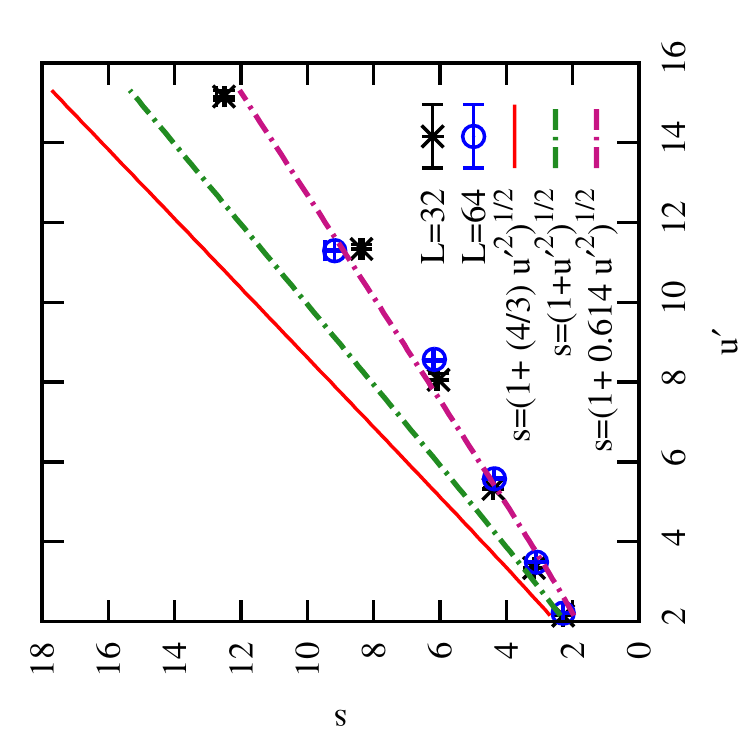}
\end{center}
\caption{Burning Velocity Diagram: Scale Invariant Models.  This figure shows a comparison between
measurements from the simulations (shown as black asterisks ($L=32$) and blue circles ($L=64$)) and
the scale invariant flame speed model $s = (1 + C_t {u'}^2)^{1/2}$, which is used as a subgrid
model in many Type Ia simulations.  The model with three different values of $C_t$ is shown: $C_t =
4/3$ (the value used in the Type Ia simulations), $C_t = 1$ (another value considered in the
formulation of the subgrid model), and $C_t = 0.614$ (a best fit to the simulation data).  The two
values associated with Type Ia subgrid models substantially overestimate the flame speed.  The best
fit model shows a pattern of residuals that indicates that it should not be extended to higher
$u'$.} 
\label{fig-turbmodels}
\end{figure}

\begin{figure*}
\begin{center}
\includegraphics[height=6in,angle=270]{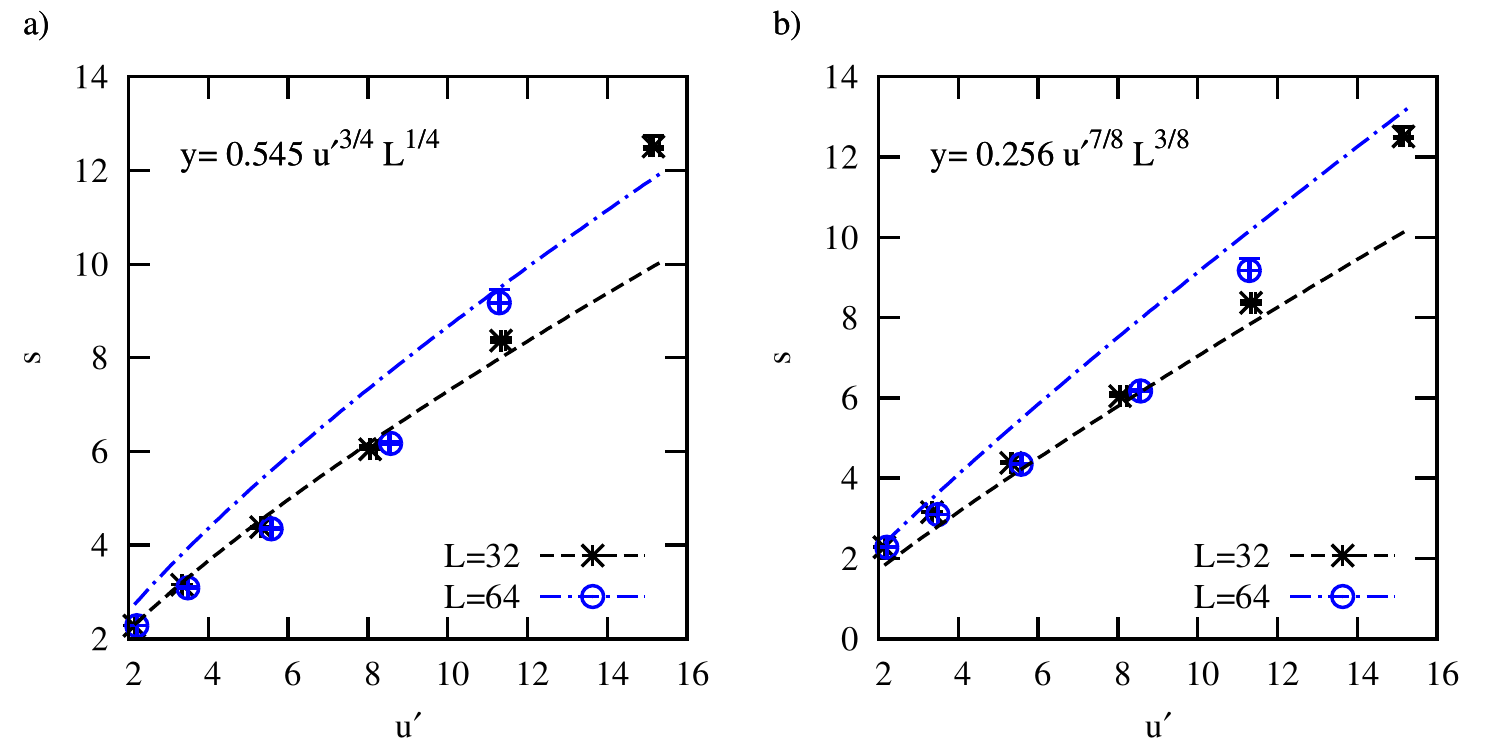}
\end{center}
\caption{Burning Velocity Diagram: Bending Models. These plots compare the simulation data ($L=32$,
black asterisks; $L=64$, blue circles) and two models from traditional turbulent combustion theory:
(a) the Zimont Model, $s = C u'^{3/4} L^{1/4}$ with best fit $C=0.545$ and (b) the Kerstein Pair
Exchange Model, $s = C_1 {u'}^{7/8} L^{3/8}$ with best fit $C_1 = 0.256$.  Both models depend on $L$
so two curves are shown for each model: the $L=32$ model curve is shown in black and should fit the
black asterisk points; the $L=64$ model curve is shown in blue and should fit the blue circle
points. Neither model adequately fits the data because the data are concave-up while the models are
concave-down.}
\label{fig-bendingmods}
\end{figure*}

\begin{figure}
\begin{center}
\includegraphics[height=3in,angle=270]{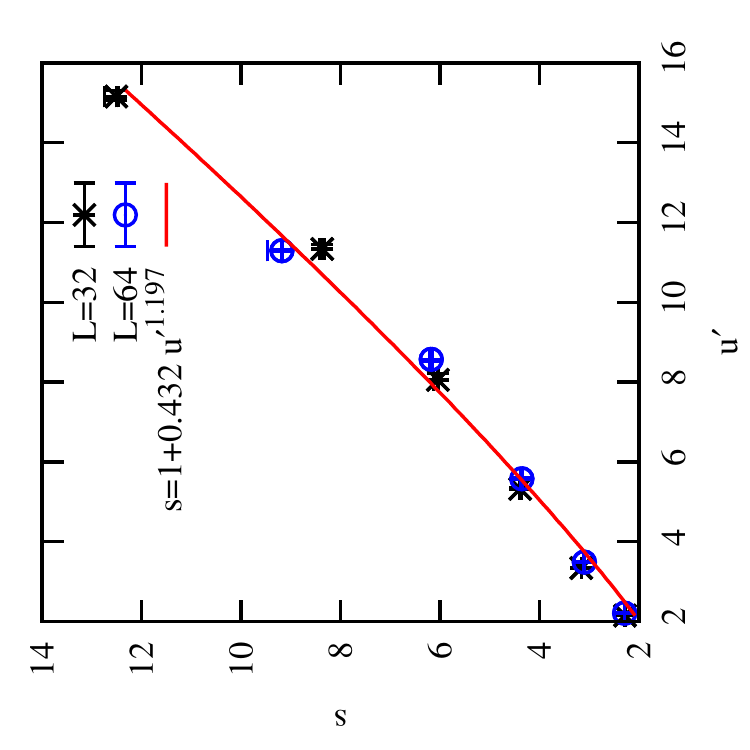}
\end{center}
\caption{Burning Velocity Diagram: Power Law Model.  This plot compares the
simulation data and a power law model $s = 1 + C {u'}^n$ with a least-squares best fit of
$C=0.432$ and $n=1.197$.  This fit shows that the simulation data are concave-up ($n > 1$), implying
that Rayleigh-Taylor unstable flames are fundamentally different from traditional turbulent
combustion flames, which show a linear or concave-down dependence ($n \leq 1$) on the same plot
axes. }
\label{fig-powerlaw}
\end{figure}

We first compare with our data with a simple linear model.  The simplest and most obvious choice of all linear models is $s =
u'$, which is the high $u'$ limit of the Damk\"{o}hler law, Equation \eqref{eqn:Dam}. A comparison between $s=u'$ and the
data from the simulations is shown in Figure \ref{fig-linearmodels} on a burning velocity diagram.  The prediction $s=u'$ is
shown as a red line and the individual measurements from the simulations are shown as black asterisks ($L=32$ simulation data
points) and blue circles ($L=64$ simulation data points).  Each simulation is represented by one point on the plot.  It is
clear that the model $s=u'$ overestimates the value of $s$. The second linear model is the full Damk\"{o}hler law, $ s= 1 + C
u'$, with a fit for the value of C. The best least-squares fit for this law (with $C=0.615$) is show in Figure
\ref{fig-linearmodels} as a dashed green line.  The model line fits the data well for the smaller values of $u'$, but
underestimates the flame speed for larger values.  The final linear model that we consider is $s=u_y'$, which is shown in
Figure \ref{fig-urmsy-model}.  Interestingly, $s=u_y'$ is a good fit for the data at low values of $u_y'$ with no fitting
parameter, although this is only true for $u_y'$ defined by averaging over the entire flame brush;  the
actual value of $u_y'$ varies considerably with height.  The fact that $s=u_y'$ fits the data well at low $u_y'$  is
consistent with the many models from traditional combustion theory in which the flame speed depends on the rms velocity in
the streamwise direction only (the $y$-direction for these simulations). However, again, the model does not fit the data well
for larger values of $u_y'$: the flame speed grows faster than a linear function of $u_y'$. Overall, linear models do not
capture the overall trend of the data in the $s$-$u'$ plane.

Second, we compare the data with scale invariant flame speed models of the type $s = (1 + C_t
{u'}^2)^{1/2}$.  This is the flame speed model used by \citet{schmidt2005, schmidt2006a,
schmidt2006b} as a subgrid model for full-star Type Ia simulations.  Figure \ref{fig-turbmodels}
shows a comparison between this model and the simulation measurements for the two values of $C_t$
used by \citet{schmidt2005, schmidt2006a, schmidt2006b}, $C_t = 4/3$ (solid red line) and $C_t =
1$ (dashed green line).  It is clear that both of these models substantially overestimate the actual
flame speed and that this overestimation is worse for intermediate values of $u'$ and then improves
slightly for large values of $u'$.  In practice, this means that Type Ia simulations using this
subgrid model may be substantially overestimating subgrid deflagration speeds.  This is not
surprising because there is no particular physical reason to expect that $C_t$ should be $1$ or
$4/3$.  In fact, \citet{schmidt2006b} suggested that $C_t$ should be a fitted parameter.  Following
this suggestion, Figure \ref{fig-turbmodels} also shows the least-squares best fit which is $C_t =
0.614$ (purple line).  This result fits the data well, but an examination of the residuals shows
that the model consistently underestimates the flame speed at low values of $u'$, overestimates it
at intermediate $u'$, and underestimates it at high $u'$. Because of this clear pattern, which is
the result of fitting an almost straight line to a curved data set, we have no confidence in an
extension of this best fit model to higher values of $u'$.  Overall, scale-invariant flame speed
models do not fit the simulation measurements well; in addition, the values of $C_t$ used in Type Ia
subgrid models significantly overestimate the flame speed.

Finally, we compare the simulation results to two models that reproduce the bending phenomena seen
in terrestrial flames.  The Type Ia subgrid model used by \citet{jackson2014} is also meant to
reproduce bending, but we will not test that model directly because of its more difficult
formulation.  To check whether models that reproduce bending fit the data well, we consider two of
the models shown by \citet{lipatnikov2002} to best fit the terrestrial flame experiments: the
Kerstein pair-exchange model \citep{kerstein1986} and the Zimont model
\citep{zimont1979,zimont1998}. Both models reproduce the bending behavior and both have a fittable
parameter. We consider them as general representatives of models that produce the bending and
thereby test the Jackson model implicitly. The Zimont model is based on kinematic wrinkling of the
flame by large eddies and thickening of the flame by small scale eddies and is given by $s = C u'
Da^{1/4}$ or $s = C u'^{3/4} L^{1/4}$ using $s_o = 1 $. Figure \ref{fig-bendingmods}, part (a)
compares this model with the simulation data.  The Kerstein pair-exchange relation models the
propagation of flamelets as a random exchange of the burned and unburned states of fluid elements in
the streamwise direction.  Using standard turbulence scalings, the turbulent flame speed is then
given by $s = C \sqrt{s_o u'} Re^{3/8}$, which is equivalent to $s = C_1 {u'}^{7/8} L^{3/8}$, using
$s_o = 1$.  A comparison of this model with the simulation data is shown in Figure
\ref{fig-bendingmods}, part (b). It is clear that neither model in Figure \ref{fig-bendingmods} fits
the data well. Fitting the low $u'$ data well inevitably results in the flame speed
for the simulation $G=32$,$L=32$ being significantly underestimated.  In addition, the dependence on
$L$ is problematic for both models because $L$ doesn't affect the flame speed at low values of $GL$.
 Overall, the problem is basically the same as with the linear models -- the data are concave-up,
while models that reproduce bending are concave-down.  No model that produces concave-down bending,
including these representative models and the \citet{jackson2014} model, will fit our concave-up
data.   

Overall, tests of linear models, scale invariant models and bending models show that none of these
models can successfully fit the data from our simulations.  The fundamental issue is that none of
these models fits the shape of the data on the burning velocity diagram.  In Figure
\ref{fig-powerlaw}, we show that the best fit of $s = 1 + C {u'}^n$ is $n=1.197$; our data curve is
concave-up, not linear (like the linear or scale invariant models at high $u'$) or concave-down
(like the bending models). This concave-up dependence of $s$ on $u'$ is different from
the concave-down dependence of typical turbulent flames. This suggests that Rayleigh-Taylor unstable
flames behave in a completely different way from flames moving through an upstream field of
turbulence.  We have shown that it is inappropriate use flame speed models from traditional
turbulent combustion theory for RT-unstable flames.

\subsection{Rayleigh-Taylor Flame Speed Model Comparison}
\label{ssec:rt}

\begin{figure*}
\begin{center}
\includegraphics[height=5in,angle=270]{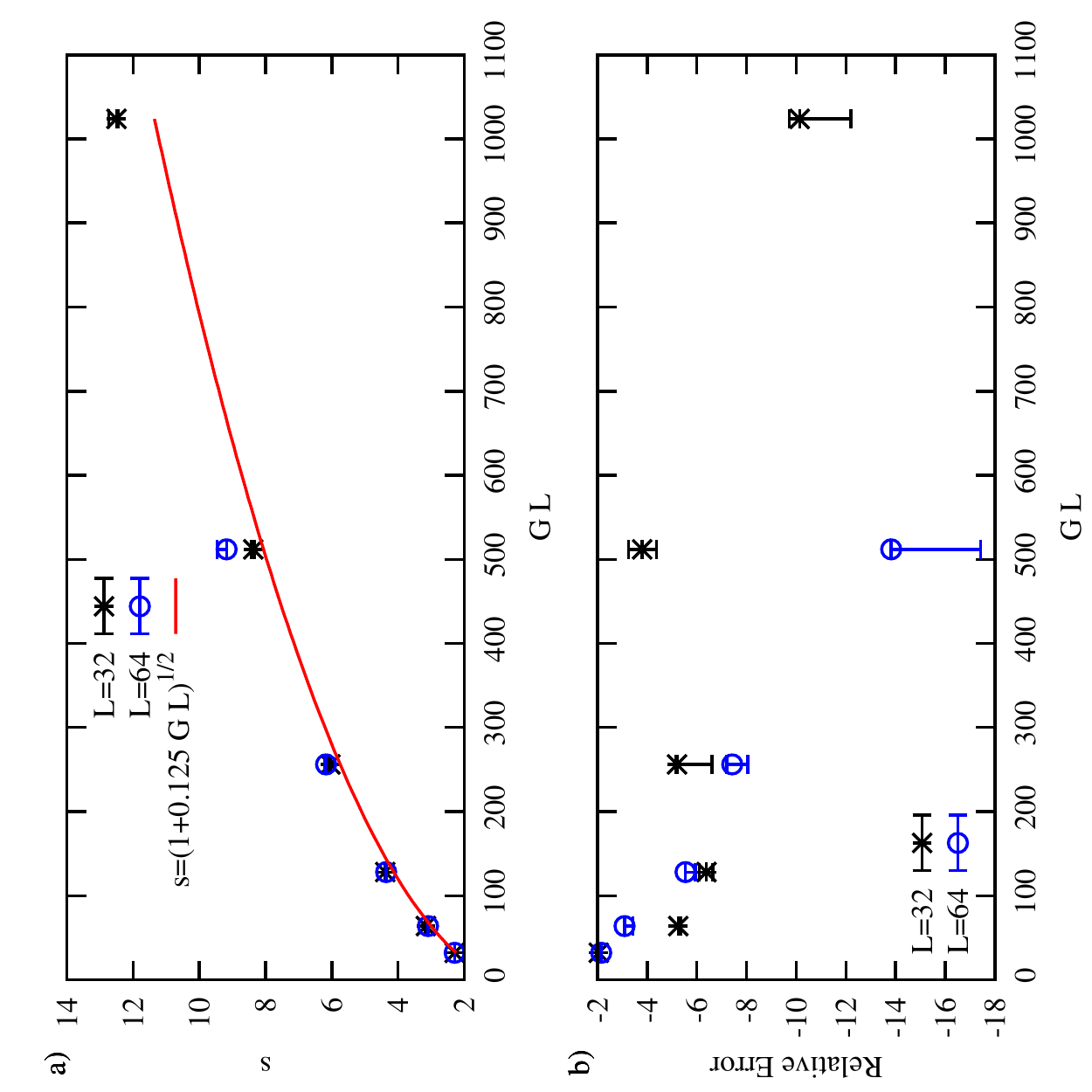}
\end{center}
\caption{Rayleigh-Taylor Flame Speed Model compared with the simulation data.  Part (a) shows a direct
comparison between the RT flame speed model prediction (solid red line) and the time-averaged flame
speeds measured from the simulations (black asterisks for $L=32$, blue circles for $L=64$).  Part (b)
shows the relative error (RE) between the predicted value (PV) and the simulation data (SD), defined
as RE=100*(PV-SD)/PV.  The error bars in both plots represent the uncertainty of averaging over a
finite-time oscillating time series (see Section \ref{ssec:measure}).}
\label{fig-rtpred}
\end{figure*}

The Rayleigh-Taylor subgrid model was first suggested by \citet{khokhlov1995} and then expanded by
\citet{zhang2007}.  In the RT model, the flame speed is determined by a balance between flame
surface creation by the instability and destruction by geometrical effects. The geometrical
destruction rate is set by the rate of collisions between flame sheets, which is determined by the
volume of the Rayleigh-Taylor bubble.  In this model, the RT instability sets the flame speed
because it controls the rates of both flame surface creation and destruction.  The flame speed
relation itself can be derived from dimensional analysis, from the linear growth rate of the RT
instability, or from the speed of a rising buoyant bubble. The expected flame speed is then, in our
dimensionless units, $s = s_o \sqrt{0.125 G L}$, for large $GL$ \citep{khokhlov1995}.  Because $G$
depends on $1/s_o^2$, this result implies that the turbulent flame speed should be independent of
the laminar flame speed. \citet{zhang2007} showed that this is the case for a carbon-oxygen flame.  

There have been several tests of the flame speed relation, both in 2D and in 3D.  In 2D,
\citet{V03,V05} confirmed the predicted RT scaling up to $GL=128$ for reflecting boundary
conditions, and $GL=512$ for periodic boundary conditions.  They corrected Khokhlov's prediction at
low values of $GL$, finding $s=s_o \sqrt{1+0.0486(G-G_1) L}$ where $G_1 = 8(2 \pi / L)^{1.72}$ is
the transition point between planar and cusped flames.  This correction ensures that the flame will
move at the laminar flame speed, $s_o$, when the $GL=0$.  At high $GL$, this is equivalent to
Khokhlov's prediction, but with a different constant because the measurements were carried out in
2D.  In a previous paper \citep{hicks2013}, these simulations were extended to $GL=16,384$ and
a best fit scaling of $s=s_o \sqrt{1+0.0503(G-G_1) L}$ was found, which is consistent with the Rayleigh-Taylor
model. All of these 2D tests were direct numerical simulations designed to resolve both the flame
width and the viscous scale.  \citet{zhang2007} carried out three-dimensional tests of the
Rayleigh-Taylor model and confirmed the RT scaling to within $10\%$ for $GL=400,671,1493,2786$.
However, these calculations did not fully resolve all scales, and, in particular, were unable to
resolve the Gibson scale even at their highest resolution.  An additional problem is that their
averages included very few flame speed oscillations.  In general, past studies of RT-unstable flames
have shown that their flame speeds are consistent with the RT model in both 2D and in 3D, but the
3D tests had some drawbacks.

Our 3D simulations are similar in some ways to \citet{zhang2007}, but they are fully resolved down
to the viscous scale and the average flame speed is computed from many more flame oscillation
periods (compare Figures \ref{fig-s-L32} and \ref{fig-s-L64} with \citet{zhang2007}, Figure 20).  We
also checked the scaling law over a large total range in $GL$, from $GL=32$ up to $GL=1024$.
Figure \ref{fig-rtpred}, part (a) shows a comparison between our
results and the 3D RT-predicted flame speed (with the correction to account for the laminar flame
behavior), $s = s_o \sqrt{1 + 0.125 GL}$.  The time-averaged flame speed is shown as black asterisks
for simulations with a domain width of $L=32$ and blue circles for $L=64$; the RT prediction is
shown as a red line.  For low values of $GL$, the RT prediction matches the data well but a
deviation of around $14\%$ is seen for the $L=64$, $GL=512$ case.  The $L=32$, $GL=1024$ case shows a
deviation of around $10\%$.  The relative error between the predictions and the simulation data
results are shown in Figure \ref{fig-rtpred}, part (b).  In this plot, the error bars based on the
uncertainty of  averaging over the oscillating flame speed can be clearly seen. These uncertainties
are not large enough to account for the deviation from the Rayleigh-Taylor model.  There are also
other uncertainties, but we have been unable to find one large enough to account for the difference
between the RT prediction and the data.  In addition, there seems to be a difference between the
$L=32$ and $L=64$ simulation flame speeds at $GL=512$ implying that there could be a domain-size
dependence that is not accounted for in the RT flame speed model (which depends only on the
product $GL$).  So, while the RT flame speed model predicts the flame speed well at low $GL$, at
higher $GL$ the turbulent flame speed is larger than predicted.  In the next section, we will
discuss a physical mechanism that could explain these higher than predicted flame speeds.

\subsection{Cusps -- The Missing Ingredient?}
\label{ssec:cusps}

In the previous section, we showed that our simulated flame moves faster than the RT flame speed for
higher values of $GL$.  In this section, we will discuss a mechanism that could produce these higher
than expected flame speeds -- the formation of cusps, either by turbulence or by the Rayleigh-Taylor
instability.  After giving a geometrical definition of two different types of cusps, we will
discuss three cusp formation mechanisms and the three different local flame speeds associated with
cusps.  Next, we will review results from \citet{poludnenko2011a} that explain how a local increase
in the flame speed can induce a faster global turbulent flame speed.  Then we will compare two
simulations that the RT model predicts should have the same flame speed, but don't, and discuss
circumstantial evidence in favor of one of the simulations being more affected by cusps than the
other.  Finally, we will consider possible measurements that could further clarify the role of
cusps.

\begin{figure}
\begin{center}
\includegraphics[width=4in,angle=0]{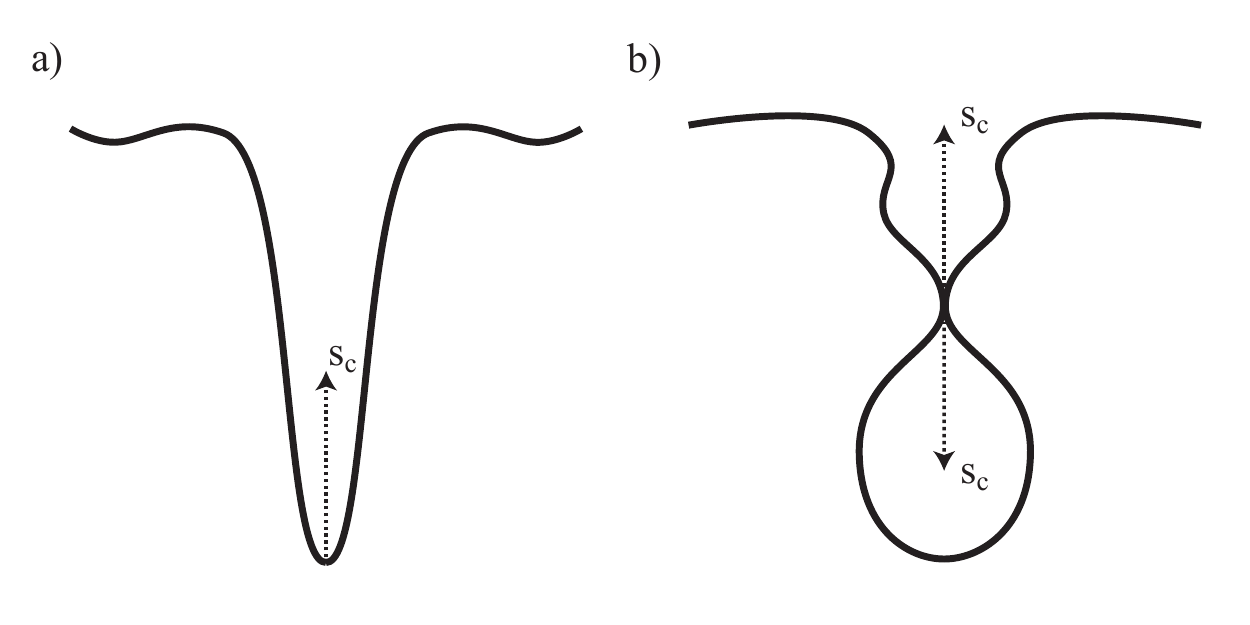}
\end{center}
\caption{Cusps: Local and Nonlocal. Part (a) shows a local cusp: a locally-formed region of high
curvature that propagates with a phase speed $s_c$.  Local cusps can be formed by the Huygens
mechanism, turbulence or the Rayleigh-Taylor instability.  Part (b) shows two nonlocal cusps that
formed when two initially distant sections of the flame surface were pushed together, either by
turbulence or by the RT instability.  This figure is based on Figure 11 of
\citet{poludnenko2011a}.}
\label{fig-cusps-lnl}
\end{figure}

\begin{figure}
\begin{center}
\includegraphics[height=4in,angle=0]{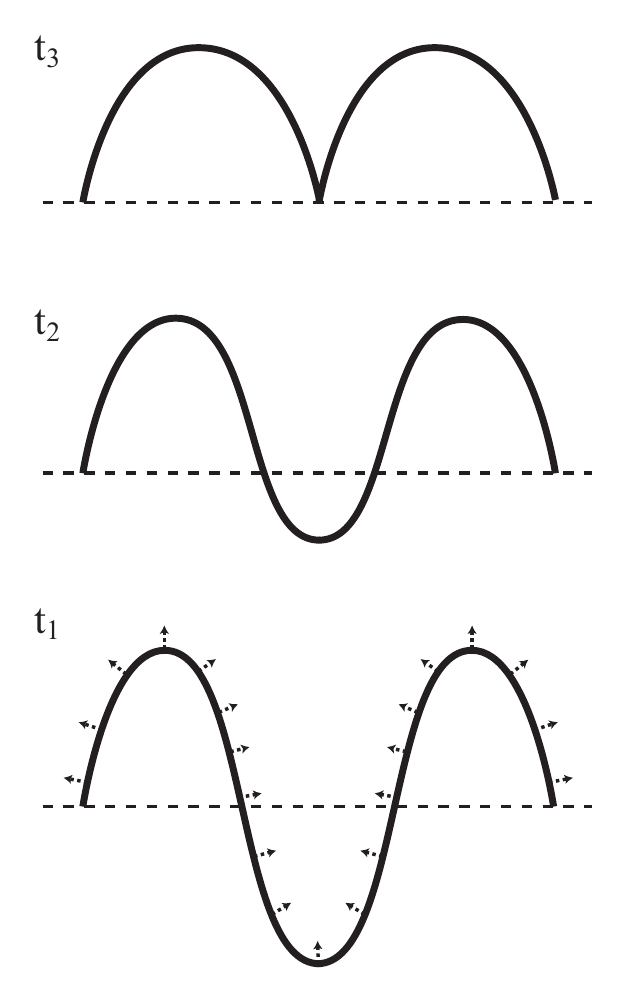}
\end{center}
\caption{Cusp formation by the Huygens mechanism.  At time $t_1$ the flame is perturbed into a
sinusoidal shape by an instability.  At every point along the flame surface, the flame propagates
normal to itself at the laminar flame speed.  At time $t_2$ this normal propagation has enhanced
the peaks of the flame surface, but is beginning to destroy the valleys. By time $t_3$ the valleys
have been completely destroyed, leaving local cusps in their place.  Now, the local normal
propagation of the flame front reinforces the cusp shape -- it is stable as long as the peaks are
not too high. This figure is based on Figure 1 from \citet{zeldovich1966}.}
\label{fig-huygens}
\end{figure}

A flame surface cusp (or ``corner'') is a part of the flame surface area where the radius of
curvature of the surface is very high (see Figure \ref{fig-cusps-lnl}).  This gives the flame
surface the appearance of a rounded v-shape with the angle between the sides of the v being small so
that the two flame sheets that comprise the sides of the v approach nearly head-on.
\citet{khokhlov1995,poludnenko2011a} have identified two different types of cusps.  A ``local'' cusp
is one that forms when a spatially-local section of a flame is deformed into a cusp shape (see
Figure \ref{fig-cusps-lnl}, part (a)).  Two ``nonlocal'' cusps form when two parametrically-distant
sections of the flame surface meet at a low angle of incidence, forming two cusps on either side of
the point of contact point (see Figure \ref{fig-cusps-lnl}, part (b)).  Once formed, local and
nonlocal cusps behave in the same way; however, their different formation mechanisms influence their
prevalence in a given flow.

A variety of physical mechanisms can trigger the formation of local and nonlocal cusps: formation
due to the normal-directional propagation of a perturbed flame surface (the Huygens mechanism), due
to turbulence or due to the Rayleigh-Taylor instability. Huygens cusp formation, illustrated in
Figure \ref{fig-huygens} occurs when a flat flame surface is slightly disturbed by a perturbation
that creates ``peaks'' and ``valleys'' on the flame surface.  Generally, each section of the flame
surface propagates normally to itself at the laminar flame speed, according to the Huygens
principle. Flame surfaces in the valleys meet, effectively creating cusps, while flame surfaces near
the peaks diverge, enhancing the peak.  The final flame surface is a series of alternating peaks and
local cusps.  \citet{zeldovich1966} first proposed this cusp formation mechanism and that the
formation of these cusps could stabilize the Landau-Darrieus instability of the flame surface.
Stabilization of this sort is not particular to the LD instability, but can also occur for flame
surfaces perturbed by the RT instability or by turbulence.

A second method of cusp formation, formation of cusps by turbulence, is extensively discussed in
\citet{khokhlov1995,poludnenko2011a}.  Turbulence can form both the local and nonlocal cusps already
mentioned. A local cusp forms when the turbulent eddies fold the flame into a cusp shape, with the
vertical extent of the ``v'' being roughly the size of the eddies.  Nonlocal cusps form when
large-scale turbulent motions fold the flame surface, pushing together sections of the flame sheet. 

\begin{figure}
\begin{center}
\includegraphics[width=3.7in,angle=0]{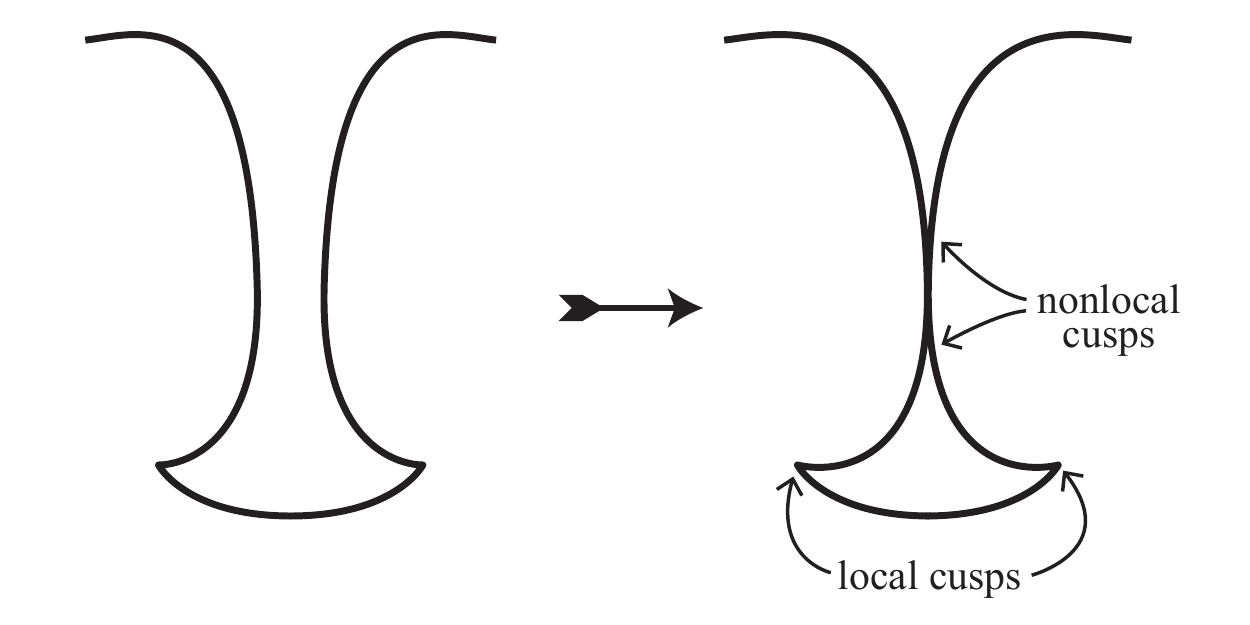}
\end{center}
\caption{Cusp Formation by the Rayleigh-Taylor Instability. This cartoon shows how the evolution of
a
Rayleigh-Taylor finger generates both local and nonlocal cusps.  As the finger evolves,
the formation of the mushroom-like head generates two local cusps (in 2D, as shown in the cartoon),
or a ring local cusp (in 3D).  As the fuel in the finger is burned, the sides of the finger approach
each other and contact, forming two nonlocal cusps.}
\label{fig-cusps-rt}
\end{figure}

\begin{figure}
\begin{center}
\includegraphics[height=3.5in,angle=0]{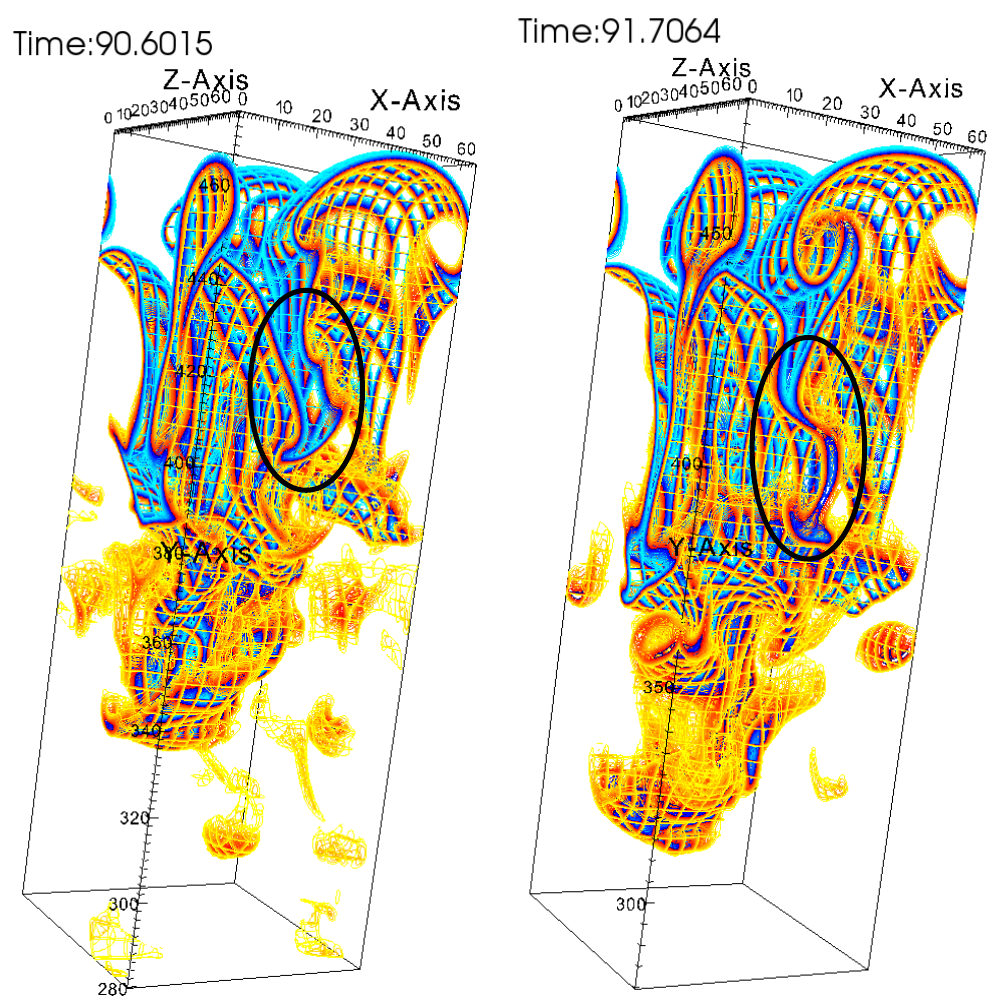}
\end{center}
\caption{Cusp Evolution.  This contour plot of temperature shows the evolution of several
RT-generated cusps in the simulation $L=64$, $G=8$.  The left side of the figure
shows a RT fuel finger (circled in black) penetrating into the burned ashes region. The right side
of the figure shows the same finger about one flame-crossing time later. The tube part of the
finger has pinched, forming two nonlocal cusps.  In addition, the local cusp in the mushroom part
of the finger has evolved, becoming sharper.  Note that several other cusps can also be seen
evolving in the figure.}
\label{fig-cusp-evol}
\end{figure}

For Rayleigh-Taylor unstable flames, we identify another mechanism of both local and
nonlocal cusp formation -- formation of cusps due to Rayleigh-Taylor fingering.  As the RT
instability acts on the flame surface it produces ``fingers'' of unburnt fuel that
penetrate into the burned ashes.  These fingers consist of a long tube of fluid connected to
a mushroom-shaped head (see Figures \ref{fig-cusps-rt} and \ref{fig-cusp-evol}). The evolution
of any individual finger creates both local and nonlocal cusps.  The 2D cartoon (Figure
\ref{fig-cusps-rt}), shows two local cusps on either side of the mushroom head. As the flame
consumes the finger of fuel, the two flame sheets that make up either side of the tube (in 2D) will
eventually come into contact forming nonlocal cusps.  In the actual 3D simulations (Figure
\ref{fig-cusp-evol}), the local cusp is a ring that wraps around the mushroom head and the nonlocal
cusps are complex structures created when parts of the tube come into contact.  For each RT finger
at least one ring-like local cusp and two or more complex non-local cusps form.  Rising bubbles of
ash can also form local cusps as they rise. Of course, this mechanism is not entirely independent of
the turbulent cusp formation mechanism, because local flows around the RT fingers do influence
their formation and evolution. 

\begin{figure}
\begin{center}
\includegraphics[height=2.3in,angle=0]{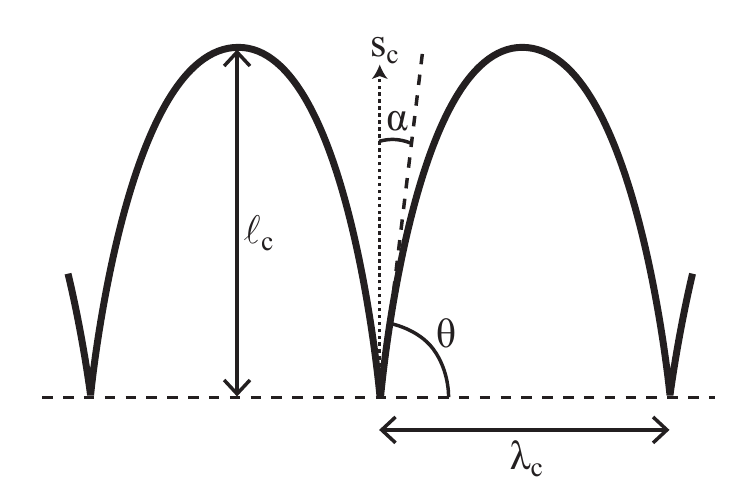}
\end{center}
\caption{Cusp Angles and Lengths. This diagrams shows various dimensions and angles associated with
a cusp.  $\alpha$ spans half of the opening angle of the cusp, $\ell_c$ is the amplitude of
the cusp and $\lambda_c$ is the wavelength of the cusp.}
\label{fig-cusp-angles}
\end{figure}

There are three basic burning speeds associated with each cusp that forms: the local laminar flame
speed ($s_o$), the cusp phase speed ($s_c$) and the local cusp burning speed ($s_n$); each of these
speeds is discussed in detail by \citet{poludnenko2011a}.  Parts of the flame far from regions of
high curvature propagate at the laminar flame speed.  The cusp itself travels at the cusp phase
velocity.  This speed is set by the geometry of the collisions between the flame sheets that make up
the sides of the cusp and does not represent a physical consumption of fuel.  The smaller the angle
($\alpha$) between the two sheets, the faster that the cusp will propagate because $s_c = s_o /
sin(\alpha)$ (see Figure \ref{fig-cusp-angles}).  This speed can be very large, and has been
measured to be as high as $s_c = 55 s_o$ for cusps studied by \citet{poludnenko2011a} and as high as
$s_c \approx 10 s_o$ for the Bunsen burner flames studied by \citet{poinsot1992}.  Thus, cusps
can disappear very rapidly after forming.  The third basic burning speed is the local cusp burning
speed, $s_n$. This speed is a magnification of fuel consumption in the cusp and is due to
the geometrical focusing of thermal diffusion in the region where the flame sheets are very close
together \citep{poludnenko2011a}.  This speed, which has its maximum, $s_n^*$, at the point of
highest curvature, depends on $\alpha$; the focusing effect is stronger at smaller angles.
\citet{poludnenko2011a} measure $s_n^* \approx 1.2 s_o$ when $\alpha = 4 \degree$ and $s_n^* \approx
3.2 s_o$ when $\alpha= 1 \degree$ (see their Figure 14, part (b)). Clearly, $\alpha$ must be very
small to significantly enhance the local cusp burning speed, but these amplifications can be
significant. The increase should be larger in 3D than in 2D.  Of the three burning speeds of the
cusp, only the local cusp burning speed reflects an actual increase in fuel consumption.  

The local cusp burning speed indicates whether the flame fuel consumption is enhanced locally, but
it doesn't quantify the extent to which the global flame speed will be amplified. Clearly, the extra
fuel consumption in the cusps will only have a large influence on the total global fuel consumption
if cusps are prevalent on the flame surface. \citet{poludnenko2011a} quantify this by showing the
total global flame speed enhancement caused by 2D cusps as a function of the total length of the
cusp, $l_c$, divided by the laminar flame width, $\delta$, for various values of $\alpha$ (see their
Figure 15). To summarize their result, global flame speed enhancements of $10 \%$ are possible for
$\alpha = 1 \degree$ for $l_c / \delta \approx 15$ and for $\alpha = 2 \degree$ for $l_c / \delta
\approx 5$. While these results can not be used to directly test whether the flame speed enhancements
observed in our simulations are due to cusp burning, it is possible to check the basic order of
magnitude for plausibility. If we assume that the RT-unstable flames in our simulations would have a
flame speed equal to the RT flame speed predicted value without the extra fuel consumption of the
cusps, then the $L=64, G=8$ simulation has approximately a $14\%$ flame speed enhancement due to
cusps and the $L=32, G=32$ simulation has a $10 \%$ enhancement due to cusps (see the model errors
in Figure \ref{fig-rtpred}). Assuming an $\alpha$ for the cusps of a few degrees, which seems
plausible because many of the cusps are non-local collisions, then if $l_c$ is relatively small
these numbers are roughly of the same order as the flame speed enhancements predicted by
\citet{poludnenko2011a}. A larger $l_c$ than those discussed by \citet{poludnenko2011a} should
produce the flame speed enhancements measured for our simulations because larger fuel consumption
amplifications are expected for 3D cusps. Overall, the theory that burning in cusps could
increase the flame speed approximately $10-15 \%$ above the RT flame speed seems plausible based on
these estimates.

Another plausibility check for the hypothesis that cusps can enhance the flame speed is to compare
two simulations that should have the same flame speed according to either the RT or turbulence-based
models yet do not.   In particular, we compare $L=32,G=16$ (Simulation A) and $L=64,G=8$ (Simulation
B).  The RT flame speed model predicts a flame speed of $s = 8.06$ for both simulations because they
each have $GL = 512$.  Turbulence-based models based on $u'$ alone also predict the same flame speed
for both simulations because they have very similar measured values of $u'_A = 11.33$ and
$u'_B=11.29$ (the $u_y'$ values are also similar).  In spite of these similarities, Simulations A
and B have significantly different flame speeds:  $s_A = 8.37$ and $s_B = 9.18$.  We will examine A
and B more closely to try to determine why their flame speeds are so different and whether this
difference is due to cusps. 

To begin, we check the flames visually for the presence of cusps.  Overall, B (Figure \ref{fig-L64-temps}, $G=8$) has more
complex cusp structures than A (Figure \ref{fig-L32-temps}, $G=16$).  Simulation A does have cusps, but they tend to be
single, large cusps instead of the many smaller cusps in Simulation B.  $L=32$, $G=32$ also has many, smaller cusps which may
explain its enhanced flame speed.  Next, we compare the flame heights of the simulations to infer the likely geometry of the
flame.  We define the flame height at any given time ($h(t)$) as the distance between the top and
bottom-most positions of any contour between $T=0.5$ and $T=0.8$. Physically, the flame height is a measure of the vertical
extent of the flame.   We averaged over time to find the average flame height, $h$, for each simulation. We found that $h_A
/L = 2.84$ but $h_B /L = 2.57$, so the flame in B is, on-average, more compact vertically than the flame in A suggesting that
B has shorter cusps than A.  Shorter cusps are consistent with a higher surface density of cusps because more cusps per
surface area means that each cusp gets less fuel and doesn't sink as far before it is burned. Both a visual inspection of A
and B and a comparison of their flame heights support the hypothesis that cusps are causing flame speed enhancements in B.

But do these cusps form mostly due to turbulence or due to the RT instability?  It is hard
to know for certain, but there are two pieces of indirect evidence that favor the dominance of RT
cusps. First, visual inspection of the flame shows that the cusps resemble evolving RT
fingers. Second, if turbulence were causing most cusp formation, simulation A should be more
affected by these cusps than B because its Karlovitz number is higher (see Figure \ref{fig-Ka}).  A
larger $Ka$ implies that the velocity at the scale of the flame width is larger for A than for B, so
A should form local turbulent cusps more effectively than B. The fact that the opposite is true -- B
seems to be more affected by cusps than A -- implies that the cusps are mostly not local cusps
formed by turbulence.  Overall, the evidence favors the cusps being formed by the RT
instability.

In this section, we reviewed and applied past work on the enhancement of the global flame speed by cusps to our simulations
of RT unstable flames.  Cusps increase the local rate of fuel consumption; the net global effect is a flame that moves faster
than expected from its surface area alone.  Our basic hypothesis is that, without cusps, the flame would move at the
RT-predicted flame speed. We suggest that any ``extra'' flame speed is due to the enhancement of the local burning velocity
by cusps. If cusps are prevalent, then these local enhancements can increase the global flame speed. To evaluate the
plausibility of this argument, we first compared the size of the speed enhancements from our simulations to predictions from
\citet{poludnenko2011a} and found that the $10-15 \%$ enhancements that we observe are not too large to be generated by
cusps. Second, we compared two simulations with different flame speeds that were predicted to have the same flame speed by
both RT and turbulence-based flame speed models. The faster of these flames visually seems to have more cusps than the slower
flame. In addition, it has a shorter normalized flame height, implying that fuel is being directed into many shorter cusps
instead of fewer longer cusps.  Finally, we argued that the cusps are more likely to have been formed by the RT instability
than by turbulence. A true understanding of the role of cusps in RT unstable flames can only be achieved by further research.
In
particular, it may be possible to directly measure the contribution of cusps by comparing the surface area of a selected
contour of the flame surface to the flame speed. \citet{poludnenko2011a} used this procedure to study turbulent flames and
showed that the flame could travel faster than accounted for by its surface area. They suggested that the excess flame speed
was due to speed amplification by cusps.  A similar analysis could be attempted for RT unstable flames. 
Moreover, direct study of the types of cusps formed by RT unstable flames, especially when paired with a cusp detection
algorithm could allow for direct measurements of cusps and detailed quantification of their effect on the flame speed.

\section{Discussion and Conclusions}
\label{sec:theend}

In this paper, we explored the physics of RT unstable flames and considered the effects of the RT
instability and the turbulence generated by the instability on the flame front.  In particular, we
considered two different types of turbulent flame speed models -- turbulence or RT based -- and
assessed their physical appropriateness and ability to fit our flame speed data. Turbulence based
flame speed models are based on traditional turbulent combustion theory, but can this theory make
successful predictions about RT unstable flames?  Does turbulence have the same effect whether it is
upstream or downstream of the flame front?  Can turbulence downstream of the flame front overwhelm
the RT stretching of the flame and set the flame speed?  The answers to these questions are
necessary to determine whether subgrid models based on traditional combustion theory should be used
in Type Ia supernovae full-star simulations. We attempted provide them in two different ways:
first, we checked whether the flame followed the traditional turbulent combustion regimes;
second, we checked the predictions of several types of turbulence-based model and the RT-based
model.  We simulated 11 different parameter combinations of $G$ and $L$ in order to probe the full
range of flame behavior from simple to complex.   To test predictions, we calculated the average
flame width, the average flame speed, the average flame height and the average rms velocity of the
turbulence behind the flame for each simulation.

To begin, we tested a basic prediction of traditional turbulent combustion theory:
the flame should transition from flamelets to reaction zones at $Ka=1$ and thicken when $Ka >1$. 
Because most of the simulations should be in the thin reaction zones regime based on their
Karlovitz number, we expected to find that the flames were thickened by eddies smaller than the
flame width.  In fact, we found just the opposite; in general, the flames were thinner than the
laminar flame width.  This suggests that the traditional picture of viscous eddies thickening the
flame front does not hold for RT unstable flames.  On the contrary, the thinning of the flame by RT
stretching seems to overwhelm turbulent thickening.  If the vorticity generated by the flame front
is rapidly washed downstream, it will not have the opportunity to interact with the flame front and
thicken it. Then RT stretching will be the main influence on
the flame front, which will be thinner than expected.  In summary, we found that the flame
did not undergo a basic regime change predicted by traditional turbulent combustion theory
and, therefore, that flame speed predictions based on this theory must be suspect.  There is
another implication of the failure to find the transition at $Ka=1$: the reduced plausibility of
the Zel'dovich gradient mechanism for DDT for these flames. The Zel'dovich gradient mechanism for
the DDT specifically depends on a large specially conditioned reactivity gradient being established
in the white dwarf.  It has been generally assumed that this gradient is produced by the thickening
of the flame as it transitions into the reaction zones regime.  However, the fact that we have
been unable to find such a transition for RT unstable flames casts doubt on this assumption. 
An important question is whether the flame will actually undergo the transition and thicken if
the turbulence is stronger than in these simulations; this is an important avenue for future
research.

The second test of whether traditional turbulent combustion theory can predict the behavior of RT unstable flames was to
compare its flame speed predictions with measurements from simulations.  We tested three basic flame speed predictions:
linear, scale invariant, and power law models.  All three types of model have been used as subgrid models for full-star Type
Ia supernovae simulations. Of the three linear models tested, $s=u'$ substantially overestimated the flame speed, and $s = 1
+ C u'$ and $s=u_y'$ underestimated the flame speed at high $u'$. It is worth noting that $s=u_y'$ did fit the data well at
lower values of $u'$ without any sort of fitting constant,  although this may be due to ``lucky
averaging'' over the variation of $u_y'$ with height.  Next, we tested the scale invariant model given by Equation
\eqref{eqn:SI} and showed that the commonly used values of $C_t = 4/3$ and $C_t = 1$ substantially overestimate the flame
speed. This suggests that the many full-star simulations using this subgrid model have flames that propagate too quickly.  In
addition, we pointed out that the scale invariant model is not physically appropriate for a RT unstable flame, since its
formulation is only valid for hydrodynamically stable flames.  Finally, we tested models meant to reproduce the ``bending
phenomena'' seen in terrestrial flames at higher values of $u'$. This sort of model has been recently adapted by
\citet{jackson2014} for astrophysical flames.  These models fail to fit our data, because the flame starts to move relatively
faster at high values of $u'$ instead of slower.  In other words, our flame speed data do bend, but they bend \textit{up}
instead of down. The fact that the flame speed curve is concave-up on the burning-velocity diagram shows why none of the
models adapted from traditional turbulent combustion theory work -- all of those models are either linear, nearly linear or
concave-down.  RT unstable flames behave in a very different way from typical turbulent flames, especially when the
turbulence is strong.  This suggests that the practice of assuming that flame speed models from traditional turbulent
combustion theory apply to RT unstable flames should be questioned.

In many ways, it is not surprising that predicting RT flame behavior from the first principles of
turbulent combustion theory should be so difficult.  Historically, it has been hard to predict the
turbulent flame speed of terrestrial turbulent flames from first principles; generally it has only
been possible to identify some very basic trends.  The turbulent flame speed often
seems to depend on factors in addition to $u'$, so developing general predictions that hold in all
regimes has been impossible so far.  In view of these difficulties, it is not surprising that
the added complication of the RT instability does not make this task easier. Overall, the
difficulties faced in predicting turbulent flame speeds in traditional turbulent combustion show how
necessary it is to rigorously test any similar model applied to astrophysical flames.

Next, we tested the predictions of the Rayleigh-Taylor based flame speed model.  This model
fit the flame speed data accurately at low values of $GL$, but underpredicted the flame speed at
higher values of $GL$.  In addition, there seems to be a dependence of the flame speed on $L$ at
$GL=512$.  If this is generally true at high $GL$, the exclusive dependence of the RT-SGS model
on the product $GL$ breaks.  This could imply a dependence of the turbulent flame
speed on the laminar flame speed, depending on how the flame speed scales with $L$.  In
summary, while the RT-SGS model predicts the flame speed at lower values of $GL$ well, it
underestimates the flame speed at higher values of $GL$.

Overall, we have shown in this paper that none of the subgrid models currently in use correctly
predict the flame speed for all values in the parameter space defined by $G$ and $L$.  Our
hypothesis is that this is because a fundamental physical phenomena is being left out of these
models -- the formation of cusps. It seems that the flame would naturally follow the RT flame speed
(since this is valid for low $GL$), except that at higher values of $GL$ the formation of cusps
enhances local burning speeds due to the geometrical focusing of thermal flux.  If cusps are
prevalent enough on the flame surface, then these local enhancements in the flame speed will affect
the global turbulent flame speed (as discussed in \citet{poludnenko2011a}).  To assess this
hypothesis, we first considered how the flame might form cusps.  Previously identified cusps
formation mechanisms are formation due to the Huygens principle \citep{zeldovich1966} and formation
due to turbulence \citep{khokhlov1995,poludnenko2011a}.  To this, we added a third mechanism:
formation due to RT fingering.  We showed that RT fingers could form both the local and nonlocal
types of cusps.  In order to gauge whether either formation by turbulence or by RT fingering could
be causing the observed flame speed excesses, we performed two basic checks.  First, we checked
whether the basic magnitude of the flame speed enhancements could reasonably be explained by cusps
and concluded that the $10-15\%$ enhancements observed were in line with the basic observations of
\citet{poludnenko2011a}.  Cusps were not ruled out by this plausibility check.  Second, we compared
two simulations with the same $GL$ and $u'$, but different flame speeds and visually identified more
small scale cusps on the faster flame.  We also showed that the faster flame took up relatively less
vertical space, again implying the presence of small scale cusps.  Finally, we suggested that the
cusps were mostly formed by the RT instability because of their visual appearance and their presence
in the larger instead of the smaller domain simulation. In summary, the formation of cusps by the RT
instability seems to be a plausible explanation for the flame speed enhancements found in this
parameter study.  This hypothesis now should be checked more rigorously and cusps and
their effect on flames should be more carefully studied.

These simulations have shed some light on the relative importance of the RT instability
and turbulence to RT unstable flames.  The dynamics of the flame seemed to
mostly be controlled by the RT instability of the flame front, not by the turbulence generated
by the RT instability.  The dominance of the RT instability leads to a thin flame that
is stretched by the RT instability instead of thickened by turbulence.  RT control of the flame also
leads to a general failure of flame speed models derived from traditional turbulent combustion
theory.  RT unstable flames are fundamentally different than traditionally turbulent flames. This
does not mean that turbulence does not influence these flames; the flow field downstream of the
flame probably acts in concert with the RT instability to shape the flame front. However, it
seems that the RT instability controls the energy budget of the system and sets the flame
speed and $u'$, at least until cusps become important.  The formation of cusps regulates the growth
of the RT instability, keeping the flame speed from growing indefinitely, but also increases the
flame speed in the process.  The formation of cusps is an avenue of flame speed enhancement that is
not directly controlled by the RT instability. 

There are several implications of these results for the choice of Type Ia subgrid model, in the
case that convective turbulence is not important (since we ignore this phenomenon).  First, the
scale invariant flame speed model
implemented by \citet{schmidt2005,schmidt2006a,schmidt2006b} is likely propagating flames too
quickly.  Second, the ``bending''-style model proposed by \citet{jackson2014} probably does not
correctly model the behavior of the flame when $u'$ is high.  As long as cusps are not important,
the RT-SGS model is a good choice.  The model $s\approx u_y'$ also predicts the flame speed well as
long as $u_y'$ is not too large.  However, based on the our results, the RT model seems to be more
appropriate physically. When cusps are important, all currently implemented subgrid models fail. 
However, cusps are probably not important to the flame unless either the RT instability or
turbulence is acting at the scale of the flame width, which does not happen until the flame
reaches the outer part of the white dwarf. For the earlier stages of evolution of the supernova,
the RT-SGS model is the best choice, but for the later stages of flame evolution, a new subgrid
model that takes cusps into account may need to be formulated.

There is a very large scope for future work on RT unstable flames, and on subgrid models for Type
Ia supernovae in general.  The research in this paper applies in the case when pre-existing
convective turbulence is not strong.  If this is not the case, then simulations are needed that
compare the relative effects of pre-existing turbulence and the RT instability on the flame speed. 
Even for the case in which convective turbulence is not strong, many avenues of exploration remain. 
It would be beneficial to explore a much wider range in the parameter space of $G$ and $L$
(especially at higher $Re$) and to vary the Prandtl number in the range $Pr < 1$, since it is
quite small in the white dwarf.  In addition, it would valuable for the simulations in this
paper to be repeated for the same parameter values but with a more realistic carbon-oxygen flame at
low Mach number, to get a sense of how all of the complicating factors ignored in these
simulations affect this simple picture.  In addition, even within the set of parameter values
studied here, a careful investigation of cusps (including attempts at direct detection of the cusps)
should be carried out to test the hypothesis that RT-generated cusps increase the flame
speed. A better understanding of cusps in general, whether formed in RT or by turbulence, is clearly
needed.

\section{Acknowledgments}
Thank you to R. Rosner for originally introducing me to Rayleigh-Taylor unstable flames and to R. Rosner and N. Vladimirova
for continuing discussions about them.  I am very grateful to P. Fischer and A. Obabko for making Nek5000 available and for
giving advice on using the code; and to N. Vladimirova for introducing me to the code and providing scripts when I first
started working on this problem. Thank you to S. Tarzia for cartoon creation, proofreading and content suggestions.  
Finally, thank you to the anonymous referee, whose helpful and insightful comments led me to improve the original version of
this paper.  I gratefully acknowledge the support of a CIERA postdoctoral fellowship during this work.  Visualizations in
this paper were generated with VisIt \citep{visit}. VisIt is supported by the Department of Energy with funding from the
Advanced Simulation and Computing Program and the Scientific Discovery through Advanced Computing Program.  This research was
supported in part through the computational resources and staff contributions provided for the Quest high performance
computing facility at Northwestern University which is jointly supported by the Office of the Provost, the Office for
Research, and Northwestern University Information Technology.

\bibliographystyle{apj}
\bibliography{Hicks-bib}
\end{document}